\documentclass[journal=aamick,manuscript=article]{achemso}

\usepackage{chemformula} 
\usepackage[T1]{fontenc} 
\usepackage{booktabs}
\usepackage{hyperref}
\usepackage{orcidlink}
\usepackage{svg}
\usepackage[dvipsnames]{xcolor}

\usepackage{xspace}

\usepackage{multirow}
\author{Surender Kumar\orcidlink{0009-0000-3072-5633}}
\affiliation{Institut für Festk\"orpertheorie und -Optik, Friedrich-Schiller-Universit\"at Jena, 07743 Jena, Germany}
\email{surendermohinder@gmail.com}

\author{Mostafa Torkashvand\orcidlink{0000-0002-5092-2711}}
\affiliation{Institut für Festk\"orpertheorie und -Optik, Friedrich-Schiller-Universit\"at Jena, 07743 Jena, Germany}
\alsoaffiliation{Department of Chemistry, Amirkabir University of Technology, 1591634311 Tehran, Iran}
\author{Stefan Velja\orcidlink{0009-0003-1268-6273}}
\affiliation{Institut für Festk\"orpertheorie und -Optik, Friedrich-Schiller-Universit\"at Jena, 07743 Jena, Germany}
\author{Caterina Cocchi\orcidlink{0000-0002-9243-9461}}
\affiliation{Institut für Festk\"orpertheorie und -Optik, Friedrich-Schiller-Universit\"at Jena, 07743 Jena, Germany}
\altaffiliation{Abbe Center of Photonics, Friedrich-Schiller-Universit\"at Jena, 07745, Jena, Germany}
\email{caterina.cocchi@uni-jena.de}

\title{Interface Symmetry and Electrostatic Stabilization of Strain-Resilient Janus Heterobilayers for Flexible Piezotronics}

\begin{document}

\begin{tocentry}
\centering
\includegraphics[]{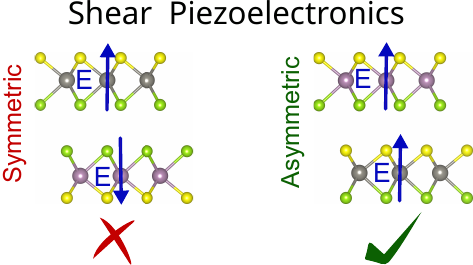}
\end{tocentry}

\newpage
\begin{abstract}
The electronic structure of conventional two-dimensional transition metal dichalcogenides (TMDs) is highly sensitive to lattice deformation, often leading to indirect-to-direct band-gap transitions that compromise performance in flexible nanoelectronic applications.  Janus TMDs, with their broken mirror symmetry and intrinsic out-of-plane dipoles, offer a promising alternative platform for electrostatic tuning. However, their electronic stability under strain and the role of the chalcogen stacking sequence in their heterostructures remains poorly understood. Here, we study from first principles the strain tolerance and piezoelectric properties of MoSSe/WSSe heterobilayers. By examining different configurations, we demonstrate that the interface chemistry strongly modulates interlayer coupling, dynamic charge redistribution, and dipole interactions. Importantly, the combined effects of intrinsic electric fields and interface electrostatics effectively suppress the strain-induced band-gap transitions typical of conventional TMDs. Moreover, while the in-plane piezoelectric response remains nearly insensitive to the stacking order, the shear piezoelectric coefficient depends heavily on the interfacial symmetry and can be effectively tuned by strain modulation. Our results highlight interfacial engineering as a powerful route to design strain-resilient Janus heterostructures for next-generation flexible optoelectronic, valleytronic, and piezotronic devices.

\end{abstract}

\newpage
\section{Introduction}
Two-dimensional (2D) transition metal dichalcogenides (TMDs), such as MoS$_2$, MoSe$_2$, WS$_2$ and WSe$_2$,  and their vertically stacked van der Waals (vdW) heterostructures are rising as  top  candidates for state-of-the-art flexible electronics.\cite{Salvatore2013,Akinwande2014,Sarkar2014,Singh2019,Lemme2020,Daus2021,Jiang2022,zheng2022,aftab2023,Katiyar2024,Yin2024,Velja2024}. Their primary appeal in this field emerges from their band gaps in the visible range, atomically smooth, dangling-bond-free surfaces, and easy interface matching to form heterostructures\cite{Gomez2022}. Their structural flexibility allows for the precise control of optoelectronic properties through layer composition and interfacial alignment for better (opto)electronics.\cite{lee2014,cheng2014,rivera2015,XIA2017,wu2019,Fan2022}

Strain engineering is another key parameter for tuning the performance of  2D TMD-based devices, allowing for unprecedented control over device performance and optical response\cite{Peng2020,Shen2016,Gomez2013,Michail2024,Velja2024}. Conventional TMDs can withstand substantial mechanical deformation without physical fracture\cite{Bertolazzi2011}. However, their electronic structure properties are rather sensitive to strain: specifically, even modest deformations on the order of 1–2\% of the equilibrium lattice parameter can modify the band character\cite{Scalise2012,Island2016,Wang2015,Zhu2013,Ramzan2025,Blundo2020}, inevitably degrading the (opto)electronic performance.  For flexible devices, this gap-character modulation represents a fundamental issue\cite{Blundo2020}, as device performance often depends on maintaining a stable electronic structure alongside high optical efficiency.

Janus TMDs, characterized by two different chalcogen species bound to the central metallic layer\cite{Zhang2017,lu2017, Dong2017, Lin2020, Zheng2021}, offer a promising alternative for overcoming some of the intrinsic limitations of conventional TMDs. Their broken out-of-plane mirror symmetry reduces the space group to C$_{3v}$\cite{Wang2026}, inducing a permanent dipole moment in the normal direction. The presence of a built-in electric field affects the excitonic properties\cite{Zheng2021,Li2017} and the spin splitting of the electronic bands\cite{Hu2018,Patel2022,Nielsen2025}, while providing  an internal, additional degree of freedom for modifying charge distribution and band alignments and introducing intrinsic piezoelectric properties\cite{Blonsky2015,Dong2017}. The permanent dipole is expected to enhance the sensitivity of the system to external perturbations such as lattice deformation, effectively buffering against the strain-induced transitions that typically degrade optoelectronic performance.

The characteristics of Janus TMDs can be further tuned in their heterobilayers\cite{Nielsen2025,Yang2025,Alfurhud2025,alfurhud2024}, where the constructive or destructive interference of the individual monolayer dipoles alters the local bonding environments, the interlayer coupling strength, and the electrostatic boundary conditions. As a result, the alignment of the interfacial atoms and their specific chemical nature emerge as structural knobs to modulate the electromechanical response of these stacks. Despite this promising tunability, a comprehensive understanding of how this interface symmetry governs the coupling between mechanical deformation and electronic properties in Janus heterobilayers remains elusive. Resolving this knowledge gap is essential for designing strain-tolerant and flexible building blocks for next-generation nanoelectronic devices.  
 
In this work, we investigate from first principles the evolution of the electronic structure and piezoelectric properties of Janus TMD heterobilayers formed by MoSSe and WSSe under a wide range of compressive and tensile biaxial strain. By comparing different interfacial configurations, we demonstrate their electronic resilience and clarify that asymmetric interfaces possess a significantly enhanced piezoelectric shear response, whereas symmetric configurations exhibit an almost vanishing shear polarization due to dipole cancellation. These findings establish clear structure–property relationships linking interface chemistry with strain tolerance, providing a robust design paradigm for engineering strain-robust 2D optoelectronic, valleytronic, and piezotronic devices.


\section{Results}

\subsection{Janus Monolayers}
\begin{figure}
    \centering
    \includegraphics[width=0.5\linewidth]{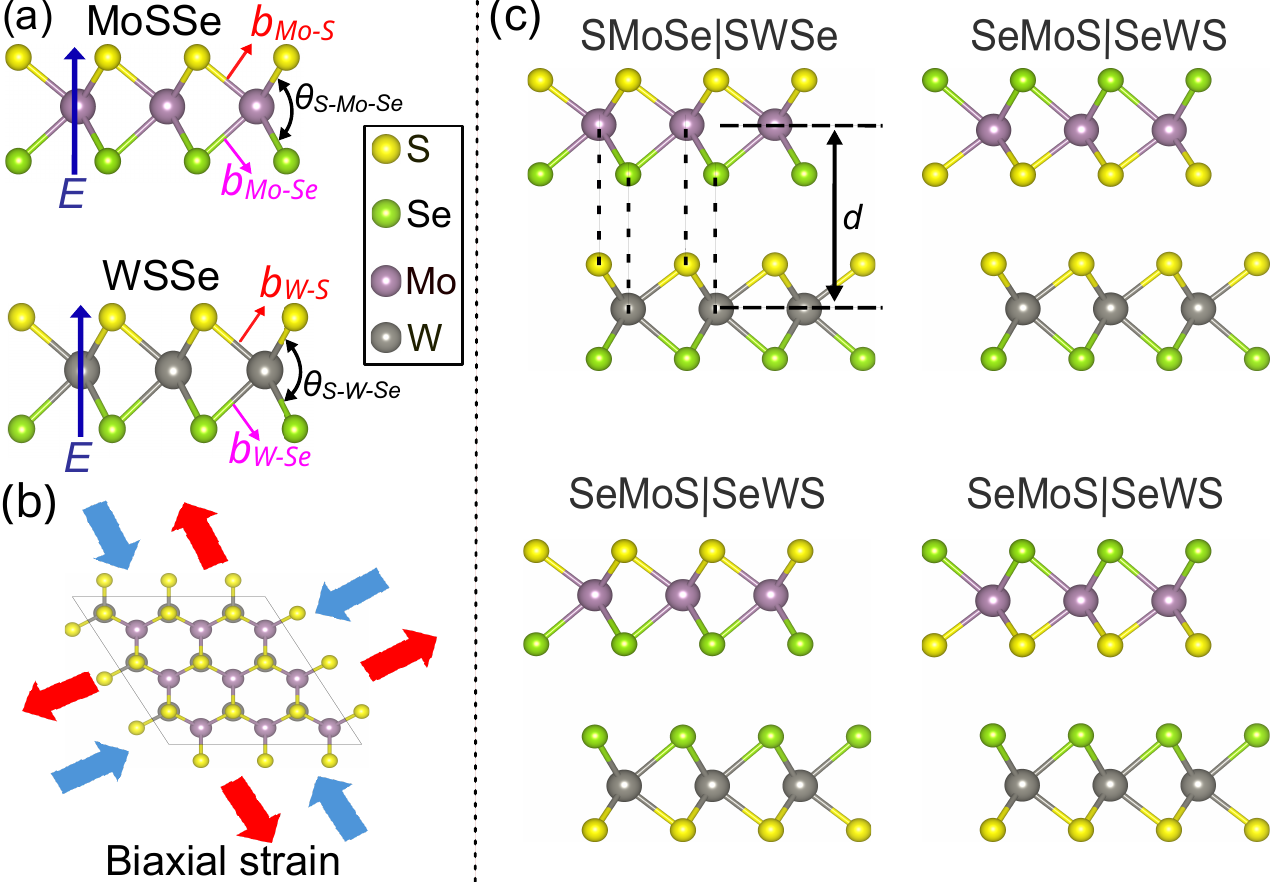}
  \caption{Side views of (a) MoSSe and WSSe monolayers, (b)  schematic illustration of the applied biaxial strain, and (c) the AB-stacked heterobilayers.. In panel (a), $b_{M\text{-}S}$ and $b_{M\text{-}Se}$ indicate the metal--chalcogen bond lengths, $\theta_{S\text{-}M\text{-}Se}$ the S--metal--Se bond angle, and $d$ the interlayer metal--metal distance, while the dark blue arrow shows the direction of the intrinsic electric field.} 
    \label{fig:structures}
\end{figure}

To establish a baseline reference, we first model Janus MoSSe and WSSe monolayers. Their primitive cell contains three atoms, with a transition metal (Mo/W) species sandwiched on either side by S and Se. This characteristic structure gives rise to an intrinsic out-of-plane structural asymmetry that induces an internal electric field $\mathbf{E}$ (Figure~\ref{fig:structures}a)\cite{Fengping2019,wei2019,Patel2022,Li2017}. After full structural relaxation (the corresponding numerical settings are reported in the Computational Section below), both monolayers exhibit an equilibrium lattice constant of $a = 3.23$~\AA{}, corresponding to the average of the in-plane lattice parameter of the MoS$_2$/WS$_2$ and MoSe$_2$/WSe$_2$ parent monolayers\cite{torkashvand2026}. Starting from these relaxed structures, we apply uniform in-plane biaxial strain (Figure~\ref{fig:structures}b), exploring the experimentally relevant range of $\pm8\%$ with steps of 2\% each. For each strain value, the lattice constant is fixed while interatomic forces are minimized again.

\begin{figure}
    \centering
    \includegraphics[width=.495\linewidth]{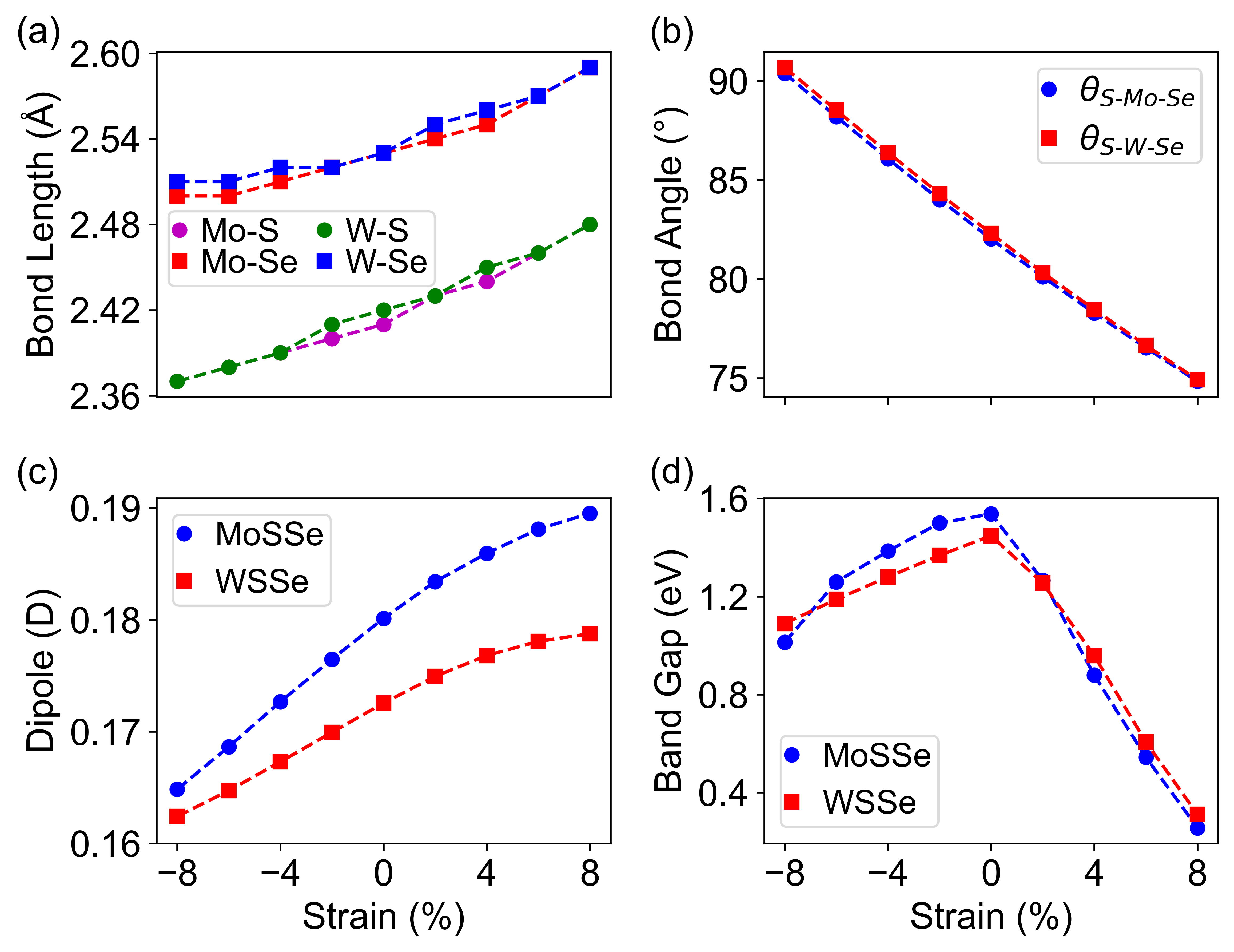}
    \caption{Strain-dependent structural and electronic properties of MoSSe and WSSe monolayers: (a) Metal--chalcogen bond lengths ($b_{M\text{-}S}$), (b) chalcogen--metal--chalcogen bond angles ($\theta_{S\text{-}M\text{-}Se}$), (c) electrostatic dipole moment, and (d) single-particle band gap.}  
    \label{fig:ml_prop}
\end{figure}

The structural and electronic properties of the MoSSe and WSSe monolayers are summarized in Figure~\ref{fig:ml_prop}. The metal–chalcogen bond lengths (Figure~\ref{fig:ml_prop}a) increase approximately linearly by $\sim$4\% from the fully compressed ($-8\%$) to the fully stretched ($+8\%$) geometry. Interestingly, the Mo–S and W–S bonds exhibit a steeper strain dependence than their Se-based counterparts, indicating a higher structural compliance that correlates with the stronger, shorter nature of the metal–sulfur coordination under deformation. The corresponding bond angles (Figure~\ref{fig:ml_prop}b) change monotonically and with strain, decreasing from $\sim90^\circ$ at $-8\%$ to $\sim75^\circ$ at $+8\%$. The near-identical behavior in MoSSe and WSSe indicates that angular changes are dominated by a universal geometric response to in-plane strain rather than transition-metal chemistry.

As both Janus monolayers are subjected to tensile (compressive) strain, the dipole moment $\mu$  increases (decreases) monotonically,  with values in MoSSe being higher than those in WSSe across the entire strain range explored (Figure~\ref{fig:ml_prop}c). Notably, $\mu$ varies by approximately 15\% over the applied deformation window, confirming that the internal electric field can be effectively tuned via mechanical manipulation. The increase in $\mu$ under tensile strain originates from a subtle competition between the stretching of polar bonds and the flattening of the monolayer, which shifts the effective centers of positive and negative charges. The band gaps of the monolayers (Figure~\ref{fig:ml_prop}d) exhibit a non-monotonic, dome-shaped dependence on strain, with the maximum value achieved around the unstrained configuration and decreasing under both compression and expansion. Interestingly, this reduction is much faster under tensile strain, dropping to $\sim$0.25 eV at +8\% increase of the lattice constant. This behavior is consistent with a strain-driven indirect gap closure reported in conventional TMDs\cite{Lei2017,Ramzan2023}. The almost identical trends found in both MoSSe and WSSe confirm that this gap-closing mechanism is a general effect governed by a valence band maximum (VBM) shift from the K-point to the $\Gamma$-point, alongside a conduction band minimum (CBM) descent at the K or $\Lambda$ valley, mirroring the well-known strain-induced band crossovers in MoS$_2$\cite{Lei2017}.  The full band structures are reported for reference in Figure~S1.

\subsection{Janus Bilayers}
\subsubsection{Structural Symmetry Breaking and Electrostatic Grading}
The heterobilayers considered in this work are constructed by vertically stacking MoSSe and WSSe in the AB-configuration, with the chalcogen atoms of one monolayer vertically aligned with the metal atom of the other sheet (Figure~\ref{fig:structures}c). The resulting four configurations, hereafter labeled SMoSe|SWSe, SeMoS|SWSe, SMoSe|SeWS, SeMoS|SeWS, are characterized by three distinct interfacial arrangements of the chalcogen species: the symmetric S|S and Se|Se interfaces and the asymmetric S|Se (or Se|S) settings.
This stacking sequence determines the orientation of the internal electric field within the bilayer\cite{torkashvand2026,Alfurhud2025}, resulting in either a constructive or destructive alignment of the constituting monolayer dipoles. 
\begin{figure}[t]
    \centering
    \includegraphics[width=.95\linewidth]{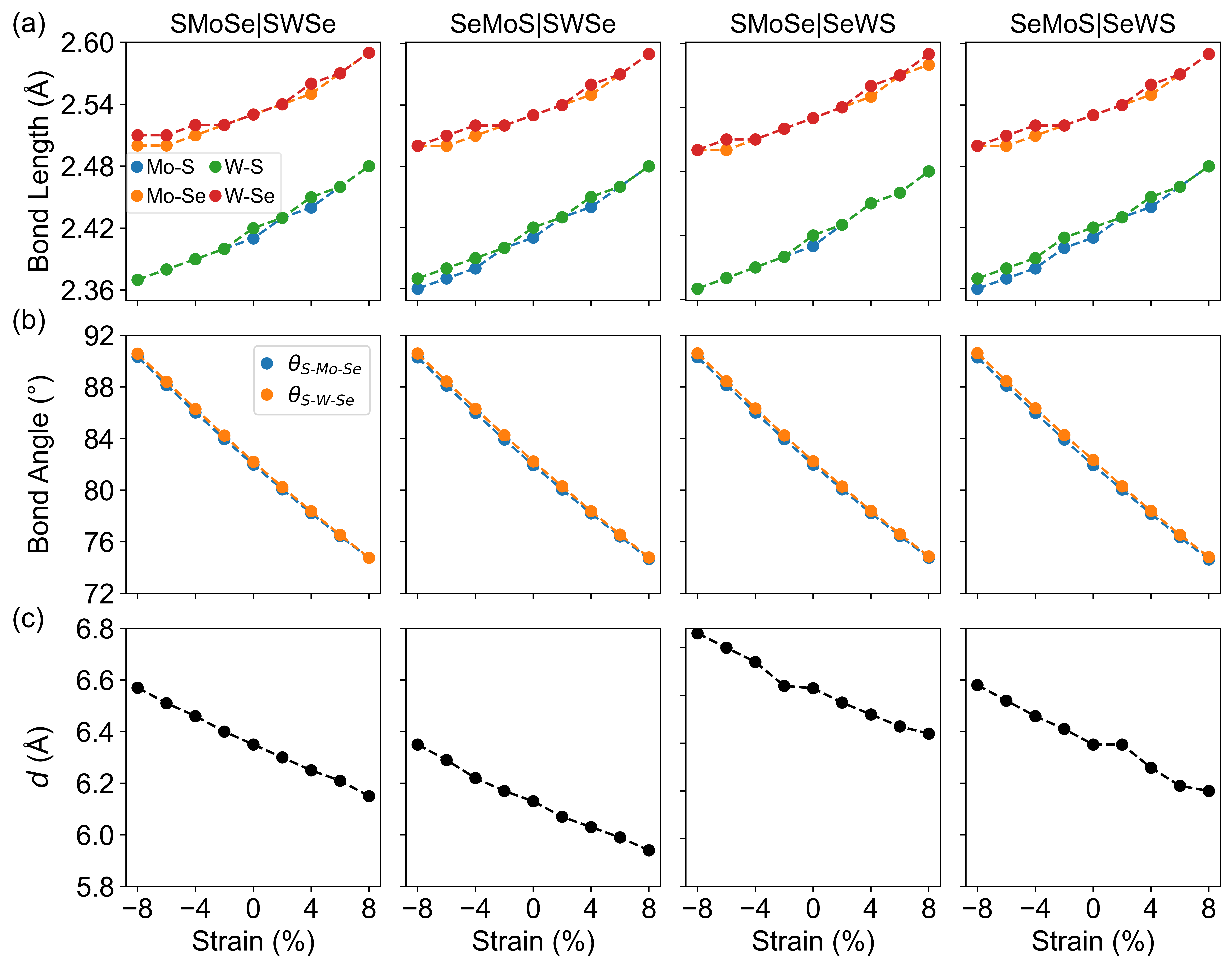}
    \caption{Strain-dependent structural properties across all considered heterobilayers: (a) metal–chalcogen bond lengths  ($b_{M\text{-}S}$), (b)  chalcogen–metal–chalcogen bond angles ($\theta_{S\text{-}M\text{-}Se}$), and (c) interlayer distances ($d$).} 
    \label{fig:BL_stru}
\end{figure} 

The bilayers are first fully optimized, and subsequently subjected to $\pm8\%$ in-plane biaxial strain. Their energetic stability is assessed through the evaluation of their formation energy ($E_f$) and interlayer binding energy ($E_b$). The former is defined as:
\begin{equation}
E_f = \frac{E_{\text{tot}} - E_M N_M - E_X N_X - E_Y N_Y}{N}.
\end{equation}
where $E_{\text{tot}}$ is the total energy of the heterobilayer, $N$ is the total number of atoms in the unit cell, and $N_i$ and $E_i$ represent the number of atoms and energy per atom of the constituent metal (M = Mo, W) and chalcogen (X, Y = S, Se) species in their respective bulk phases. To quantify the interlayer coupling strength, the interlayer binding energy is computed as:
\begin{equation}
E_b = E_{\text{tot}} - E_{\rm MoSSe} - E_{\rm WSSe},
\label{eq:E_b}
\end{equation}
where $E_{\rm MoSSe}$ and $E_{\rm WSSe}$ are the total energies of the isolated, fully relaxed Janus monolayers. The negative values of both $E_f$ and $E_b$ obtained across all configurations and strain range (Table~S1), confirm that these heterobilayers are energetically stable\cite{Mahajan2026,Wenyu2020,van2020}. 

For all considered heterobilayers, both qualitative and quantitative trends in bond lengths (Figure~\ref{fig:BL_stru}a) and bond angles (Figure~\ref{fig:BL_stru}b) closely match those obtained in the corresponding monolayers. Under tensile strain, the metal–chalcogen distances increase, while the S–M–Se angles decrease, with the opposite behavior occurring under compression. Interestingly, for a given strain value, the bond lengths and bond angles are nearly identical across all four configurations. This finding indicates that the specific interface arrangement plays a negligible role in determining the structural properties of these heterobilayers. 

Next, we explore the effect of in-plane biaxial strain on the interlayer distance \textit{d}. In the equilibrium geometry, the SeMoS|SWSe heterobilayer is characterized by the smallest separation ($d=6.2 \AA{}$), while the SMoSe|SeWS configuration shows the largest value ($d=6.6 \AA{}$), with the two mixed-interface configurations lying in between\cite{torkashvand2026} (Figure~\ref{fig:BL_stru}c). This trend is consistent with the smaller atomic radius of S compared to Se, allowing a closer vertical proximity of the monolayers, in contrast to the larger steric separation induced by the interfacial Se atoms. 
Subject to strain, the interlayer spacing exhibits an inverse response: under compression \textit{d} increases, whereas under expansion it decreases. This trend appears consistently across all four configurations and is a direct consequence of the positive out-of-plane Poisson ratio of the heterobilayers, which drives a structural expansion in the vertical direction to accommodate the in-plane lattice reduction. The linear behavior of \textit{d} with respect to strain indicates a purely elastic and reversible electromechanical response within the explored strain range. 

\begin{figure}[h]
    \centering
    \includegraphics[width=0.5\linewidth]{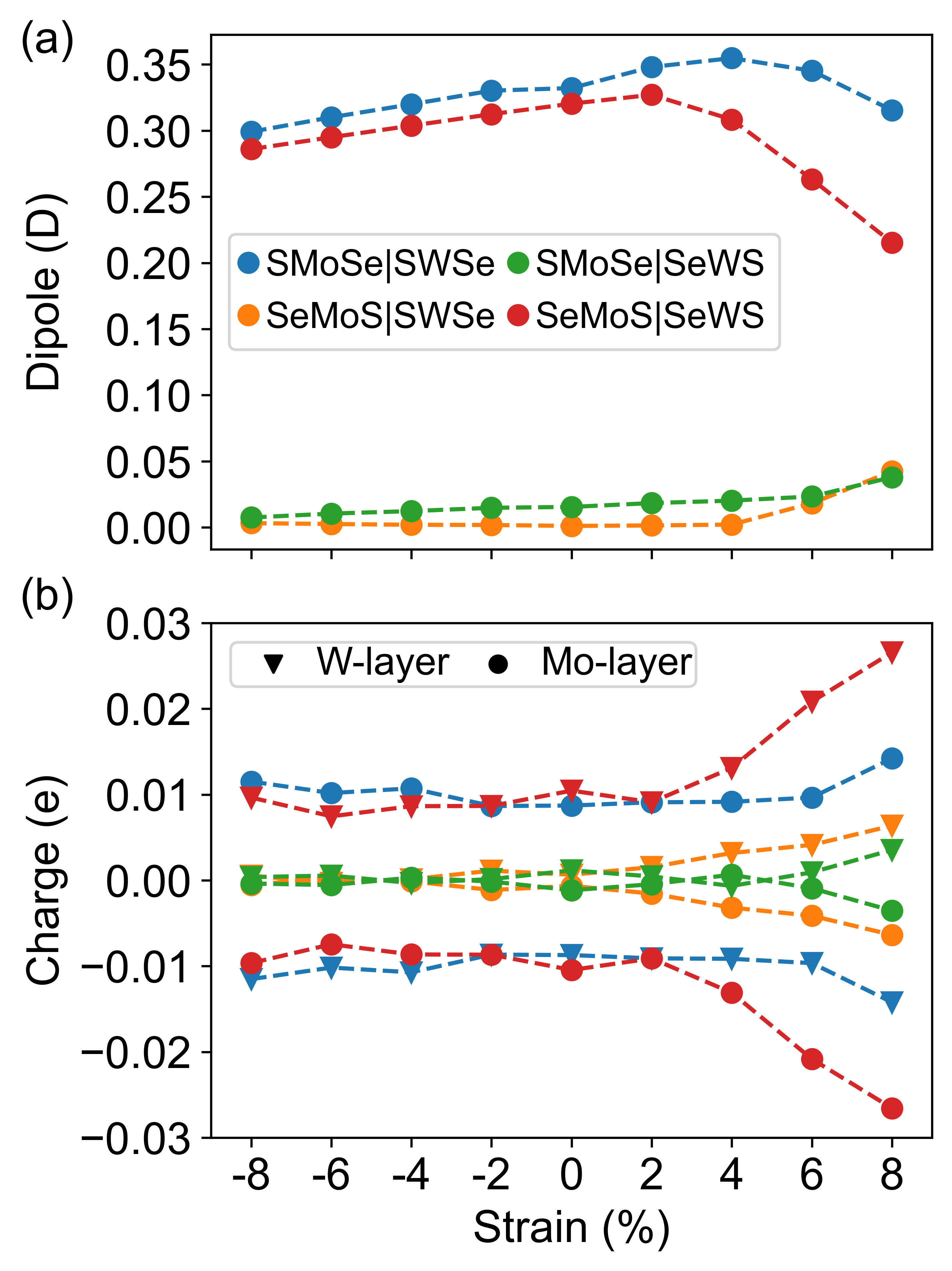}
    \caption{Variation of (a) the net internal dipole moment and (b) the monolayer-resolved Bader charges, tracked by the transition-metal species, as a function of biaxial strain in the four heterobilayers.}
    \label{fig:BL_ele}
\end{figure} 

The most striking characteristic of the heterobilayers is the variation of the net out-of-plane dipole moment as a function of strain (Figure~\ref{fig:BL_ele}a). In all four configurations, $\mu$ exhibits a non-linear, dome-shaped profile peaked at 2\% deformation. This trend strongly contrasts with the monotonic behavior in the monolayers (Figure~\ref{fig:ml_prop}c) and highlights the non-additive nature of electrostatic dipoles in stacked Janus TMDs. In the heterobilayers, the internal electric field is the result of a delicate interplay between the internal polarization of the constituent monolayers and the strain-induced modulation of the interlayer distance. Around the unstrained geometry, the system balances the internal monolayer polarizations at its equilibrium interlayer separation (Figure~\ref{fig:BL_stru}c). Under either tensile or compressive strain, the out-of-plane Poisson response alters the vertical separation, shifting the local charge transfer and the electrostatic response, ultimately leading to the observed dome-shaped behavior.

A relative comparison among the four configurations reveals specific electrostatic profiles dictated by the interface symmetry (Figure~\ref{fig:BL_ele}a). The SMoSe|SWSe stackings host parallel alignments, where the constituent monolayer dipoles add constructively, with the former achieving the highest net dipole moment ($\mu=$0.34 D). In contrast, the SeMoS|SWSe and SMoSe|SeWS features are nearly anti-parallel dipoles, leading to a substantial electrostatic cancellation across the entire strain range. The slight variations in magnitude between configurations with identical dipole orientations (such as SMoSe|SWSe vs. SeMoS|SeWS) stem from subtle differences in the local electronic charge redistribution, which depends on whether S or Se atoms  at the interface are attached to Mo or W. As a result, the stacking orientation and interface chemistry emerge as knobs to tailor the net electrostatic landscape of the heterobilayer.

The variation of \textit{d} with respect to strain (Figure~\ref{fig:BL_stru}c) induces an interlayer charge redistribution. Although small in magnitude ($\sim$0.02~e$^-$ or less, as shown in Figure~\ref{fig:BL_ele}b), it can significantly alter the resulting electronic structure. In the symmetric heterobilayers, featuring S|S and Se|Se interfaces, the net interlayer charge transfer remains near zero, showing only a slight increase under high tensile strains. Conversely, the asymmetric configurations (Se|S and S|Se interfaces) induce a significantly strain-sensitive charge transfer, especially in the large tensile-strain regime. This behavior suggests that the interface chemistry, and particularly the orbital overlap between the facing chalcogen species, is highly sensitive to changes in the interlayer distance. This finding is physically important because a strain-dependent charge transfer implies that the interfacial band alignment shifts under mechanical loading. The trends of individual atom Bader charges with strain are reported in the SI (Figure~S2).


\subsubsection{Electronic Structure, Band Alignment, and the Role of Chalcogen Interfacial Layers}\label{sec:ele}

\begin{figure}
    \centering

    \includegraphics[width=0.485\linewidth]{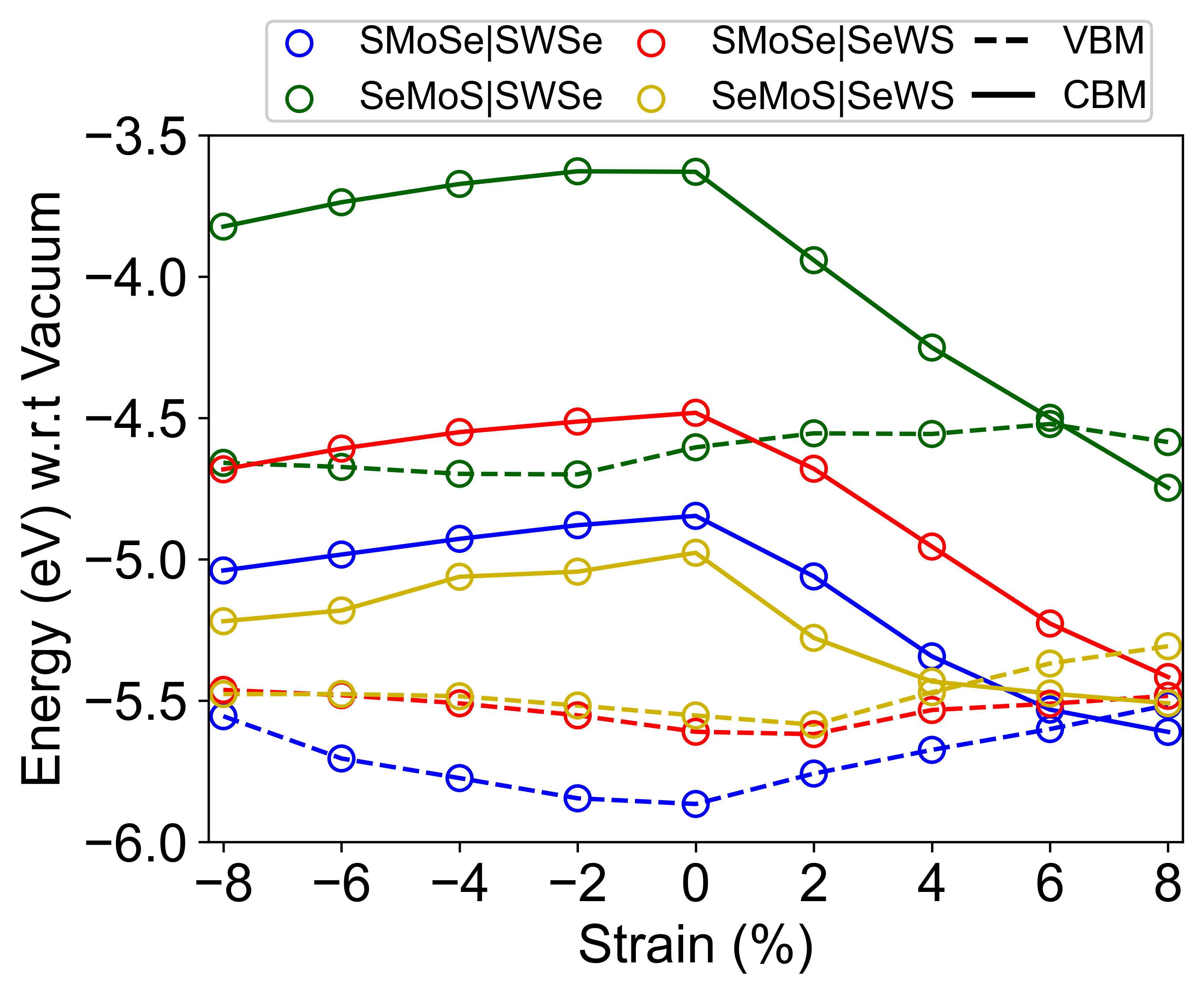}
    \caption{ Variation of absolute VBM and CBM w.r.t. vacuum level with biaxial strain for all bilayers.   }
    \label{fig:BL_offset}
\end{figure}

We continue our analysis by exploring how strain influences the electronic structure, focusing on the relative level alignments and layer-projected band structures of the heterobilayers. While the magnitude of the band gap reflects the energy difference of the frontier states, the absolute positions of the band edges are critical for predicting interfacial charge transfer efficiency, assessing material stability, and evaluating redox capabilities for photocatalytic applications\cite{torkashvand2026,Ju2020,wei2019}. The evolution of the conduction band edge under strain (Figure~\ref{fig:BL_offset}) reveals a distinct valley-reconfiguration threshold under tension, where a sharp energetic discontinuity emerges due to a band crossover between competing valleys. In contrast, the VBM exhibits remarkable resilience against lattice deformation, remaining energetically very close to its value at equilibrium, with shifts on the order of 100 meV. The SMoSe|SWSe configuration is the only exception, where the larger interlayer orbital hybridization leads to a more pronounced valence band edge shift of $\sim$300 meV. 

\begin{figure}[t]
    \centering
    \includegraphics[width=0.98\linewidth]{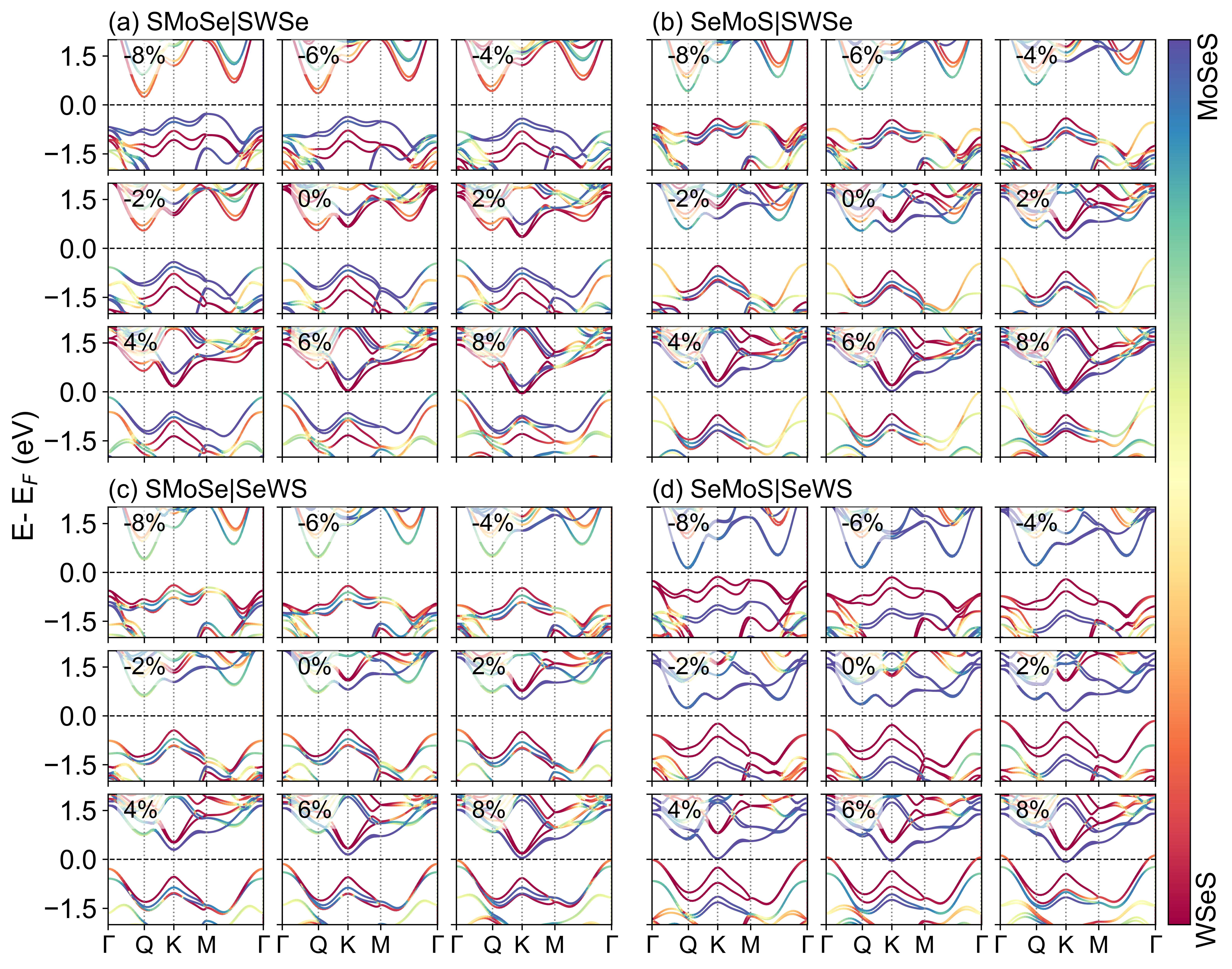}
    \caption{Strain-dependent layer-projected band structures of (a) SMoSe|SWSe, (b) SeMoS|SWSe, (c) SMoSe|SeWS, and (d) SeMoS|SeWS heterobilayers. The Fermi energy (E$_F$) is set to zero.}
    \label{fig:BL_bands}
\end{figure}

 The strain-dependent band structures, projected onto the individual monolayer contributions (Figure~\ref{fig:BL_bands}), reveal the degree of hybridization and interlayer coupling. Both features are highly sensitive to strain through the modulation of the interlayer distance, reflecting the influence of interface geometry\cite{Wang2019}. At equilibrium, all bilayers exhibit a type-II band alignment\cite{Wenyu2020,torkashvand2026, Nielsen2025}. In the SMoSe|SWSe configuration, the VBM is localized on MoSSe and the CBM on WSSe, while this layer contribution is reversed (VBM on WSSe and CBM on MoSSe) in the other three stacks. Among the four heterobilayers, SeMoS|SeWS is the only configuration hosting a direct band gap at \textit{K} (Figure~\ref{fig:BL_bands}d), while the remaining three systems are indirect-gap semiconductors. Specifically, in SMoSe|SWSe and SMoSe|SeWS, the VBM lies at \textit{K} while the CBM resides in the \textit{Q}-valley (Figure~\ref{fig:BL_bands}a,c). SeMoS|SWSe is qualitatively unique, with the VBM located at the $\Gamma$-point, approximately 500~meV higher in energy than the local maximum at \textit{K}, while the CBM occupies the \textit{Q}-valley (Figure~\ref{fig:BL_bands}b).
 
The degree of interlayer hybridization varies significantly across all configurations, even at zero strain (Figure~\ref{fig:BL_bands}). The symmetric SeMoS|SWSe bilayer exhibits the strongest mixing (Figure~\ref{fig:BL_bands}b), particularly along the $\Gamma$–\textit{K} direction in the valence region, where nearly equal contributions from both layers appear. This strong hybridization is a direct consequence of its compact S|S interfacial arrangement, which yields the shortest interlayer distance ($d = 6.2$~\AA{}, see Figure~\ref{fig:BL_stru}c) and thus maximizes out-of-plane orbital overlap.  In contrast, the other bilayers display much weaker hybridization. This behavior is especially evident in the SeMoS|SeWS configuration, where both the valence and conduction band edges are characterized by monolayer-localized states (Figure~\ref{fig:BL_bands}d). 

Biaxial strain dramatically reshapes the electronic landscape of the heterobilayers by altering the relative energies of the local valleys, triggering a competition that drives the nature of the band gap and its magnitude. Under increasing tensile strain, both the conduction and valence states shift downward. However, while  \textit{Q}-valley remains nearly rigid, varying by only $\sim$400~meV across the considered strain range, the \textit{K}-valley is highly sensitive to mechanical deformation, exhibiting variations up to 2~eV (Figures~\ref{fig:BL_bands} and S3). As a result, the K-valley shifts upward steeply under compression (-2\% to -8\%), leaving the CBM at the less sensitive \textit{Q}-valley. In contrast, tensile strain substantially lowers the \textit{K}-valley energy, making it the dominant contributor to the CBM. The conduction band edge is thus primarily governed by this strain-driven \textit{K–Q} valley competition, while states at other high-symmetry points (like \textit{M}) remain much higher in energy and play a negligible role in determining the CBM (Figures~\ref{fig:BL_bands} and S3). 

In the valence band, the strain-induced valley dynamics are highly configuration-dependent, triggering a complex competition among the \textit{K}-, $\Gamma$-, and \textit{M}-points. Under high compression, these three valleys converge in energy and begin to mutually compete (Figure~S3). On the other hand, under increasing tensile strain, their sensitivities diverge drastically: the $\Gamma$-valley remains remarkably rigid, shifting by only ~400 meV across the entire strain window, whereas the \textit{K}- and \textit{M}-valleys exhibit a substantial strain dependence, with energy shifts of about 1.5 and 2~eV, respectively. Specifically, the energy of the \textit{K}-valley decreases linearly with tensile strain, while the \textit{M}-valley falls even deeper into the valence band. Consequently, the VBM is determined by the rigidity of the $\Gamma$-valley and its interplay with the shifting \textit{K}-state under deformation.

At sufficiently large tensile strain ($\geq5\%$), the heterobilayers undergo a semiconductor-to-metal (or semimetal) transition\cite{Bhattacharyya2012}, driven by the valence band edge at $\Gamma$ crossing the conduction  \textit{K}-valley near the Fermi level. Remarkably, for all configurations except SeMoS|SeWS, the band gap remains indirect within a wide, experimentally exploitable strain window ranging from approximately $-6\%$ to $2\%$. This indirect nature stands in stark contrast to conventional homobilayers like MoS$_2$, which undergo an indirect-to-direct band-gap transition under strain. In addition, the direct band gaps at different high-symmetry points (Figure~S4) indicate that the \textit{K}-valley consistently maintains the smallest direct gap in the low-strain regime. This behavior suggests that the \textit{K-K} transition will dominate the optical spectrum and excitonic properties of these Janus heterobilayers, providing a promising avenue for future optoelectronic investigations beyond the present work.

Under in-plane compression, the interlayer hybridization is significantly reduced, causing the electronic bands to become increasingly localized on their respective monolayers (Figure~\ref{fig:BL_bands}). In contrast, tensile strain enhances state mixing between the two layers. This behavior is governed by the strain-induced modulation of the interlayer separation: in-plane compression drives an out-of-plane expansion that increases $d$, thereby weakening the interlayer orbital overlap. Conversely, tensile strain reduces the vertical separation, forcing the layers closer together and promoting stronger interlayer orbital coupling. These structural variations fundamentally dictate the overall electromechanical and electronic responses of the heterobilayers. 

\subsubsection{Piezoelectric Properties}
A technologically important consequence of the broken inversion symmetry in anisotropic materials is piezoelectricity, which establishes a direct coupling between mechanical deformation and electrical polarization. Piezoelectric materials, efficiently converting mechanical and electrical energy, are central to the design of sensors, actuators, and transducers\cite{king1990,tressler1998,dagdeviren2014,zhao2021,kim2022,He2024,jiang2024,Chen2025}. Microscopically, the direct piezoelectric effect manifests as a strain-induced electric polarization, described by a third-rank tensor $e_{ijk}$. In Voigt notation, this reduces to a second-rank tensor $e_{im}$ ($m=1, 2,…, 6$). For bilayer systems with $C_{3v}$ point-group symmetry\cite{Lin2025}, the piezoelectric tensor takes the form:
 \begin{align}
e_{im} =
\begin{pmatrix}
0 & 0 & 0 & 0 & e_{15} & -e_{22} \\
-e_{22} & e_{22} & 0 & e_{15} & 0 & 0 \\
e_{31} & e_{31} & e_{33} & 0 & 0 & 0
\end{pmatrix},
\end{align}
yielding four independent non-zero coefficients: $e_{22}$, $e_{31}$,$e_{33}$ and $e_{15}$. Specifically, $e_{22}$ describes the in-plane polarization induced by uniaxial in-plane strain, $e_{31}$ and $e_{33}$  correspond to out-of-plane polarization generated by strain along the $x$- and $z$-directions, respectively, and $e_{15}$  characterizes the shear-induced polarization.

\begin{table}[ht]
\centering
\caption{Calculated piezoelectric tensor components (in C/m$^2$) for the considered Janus TMD heterobilayers under biaxial strain ranging from $-6\%$ to $+2\%$, presented in Voigt notation.}
\begin{tabular}{c c c c c c}
\hline
Structure & Strain (\%) & $e_{22}$ & $e_{31}$ & $e_{33}$ & $e_{15}$ \\
\hline

\multirow{4}{*}{SMoSe|SWSe}
&-6 & 0.025 & 0.017&0.010 & 0.122\\
 & -4 & 0.026 & 0.017 & 0.012 & 0.113 \\
 & -2 & 0.027 &0.017 & 0.013 & 0.108 \\
 &  0 & 0.028 & 0.017 & 0.014 & 0.105 \\
 &  2 & 0.028 & 0.017 & 0.015 & 0.104 \\

\hline
\multirow{4}{*}{SeMoS|SWSe}
 & -6 & 0.024 & 0.004 & 0.004 & 0.000 \\
 & -4 & 0.025 & 0.003    & 0.002 & 0.001 \\
 & -2 & 0.026 & 0.005 & 0.004 & 0.002\\
 &  0 & 0.027 & 0.001 & 0.000 & 0.003 \\
 &  2 & 0.027 & 0.002 & 0.000 & 0.004 \\

\hline
\multirow{4}{*}{SMoSe|SeWS}
 & -6 & 0.024 & 0.007 & 0.006 & 0.002 \\
 & -4 & 0.025 & 0.011 & 0.009 & 0.001 \\
 & -2 & 0.026 & 0.007 & 0.006 & 0.000\\
 &  0 & 0.027 & 0.013 & 0.012 &0.001 \\
 &  2 & 0.028 & 0.007 & 0.006 & 0.001 \\

\hline
\multirow{4}{*}{SeMoS|SeWS}
 & -6 & 0.024 & 0.014 & 0.007 & 0.124 \\
 & -4 & 0.025 & 0.014 & 0.009 & 0.116 \\
 & -2 & 0.026 & 0.014 & 0.011 & 0.110 \\
 &  0 & 0.027 & 0.013 & 0.012 & 0.107 \\
  &  2 & 0.027 & 0.012 & 0.013 & 0.107 \\

\hline
\end{tabular}\label{tab:piezo}
\end{table}

The calculated piezoelectric coefficients across the considered strain window (Table~\ref{tab:piezo}) reveal that $e_{22}$ remains nearly insensitive to both external strain and stacking configuration, varying only within a 0.024–0.028 C/m$^2$ range. The marginal reduction under compression and the subtle enhancement under tension indicate that the in-plane piezoelectric response is mainly ruled by the intrinsic hexagonal lattice symmetry of the constituent TMD Janus monolayers, rather than by interlayer electrostatic variations. Similarly, the out-of-plane coefficients $e_{31}$ and $e_{33}$ show a minimal strain dependence and remain consistently small across all structures. 

Strikingly, the most pronounced structural sensitivity is captured by the shear coefficient $e_{15}$, which effectively splits the heterobilayers into two functional regimes: shear-active and shear-suppressed systems. In configurations featuring parallel dipole alignments, such as SeMoS|SeWS and SMoSe|SWSe, the shear response is exceptionally large ($e_{15} \sim 0.11$~C/m$^2$), while configurations with antiparallel dipole alignments (SeMoS|SWSe and SMoSe|SeWS) exhibit a nearly vanishing shear response. This behavior highlights the decisive role of stacking-induced macroscopic inversion symmetry in regulating shear polarization. In antiparallel stackings, the opposing intrinsic monolayer dipoles cancel each other, effectively suppressing shear-induced electronic displacement. In parallel stackings, on the other hand, the constructive reinforcement of the monolayer static dipoles breaks this macroscopic symmetry, unlocking a sizeable shear response. These results demonstrate that shear piezoelectricity in Janus TMD bilayers can be effectively engineered through symmetry-controlled stacking design.

\subsubsection{ Born Effective Charges}
Born effective charges (BEC) provide a microscopic measure of electromechanical coupling by quantifying the change in macroscopic polarization induced by infinitesimal atomic displacements. Formally, the BEC tensor component is defined as~\cite{Yumnam2018}:
\begin{equation}
    Z^*_{ij}=\Omega\frac{\partial P_i}{\partial r_{j}},
\end{equation}
where $\Omega$ is the unit-cell volume and $P_i$ is the component of the polarization vector along direction $i$ generated by the displacement of an atom along $j$. Large anomalous BECs are characteristic of materials with strong charge redistribution under ionic motion and are closely related to piezoelectric and ferroelectric responses.\cite{Alam2019,Choudhary2020,Hasin2024,Hong2024}

\begin{figure}
    \centering
    \includegraphics[width=0.5\linewidth]{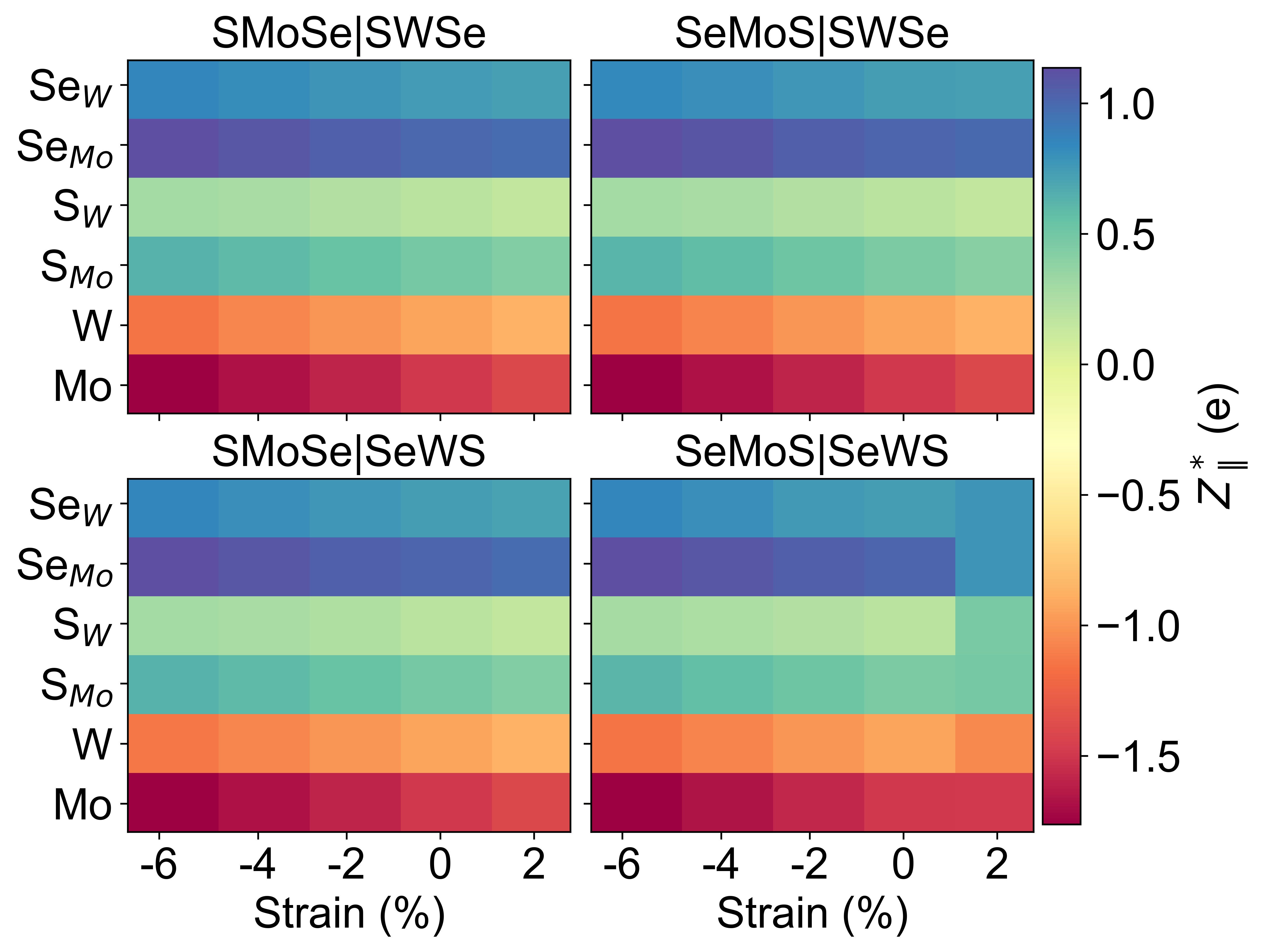}
    \caption{Evolution of the in-plane components of the Born effective charges (Z$^*_{\parallel}$) for individual atoms as a function of applied biaxial strain across the four considered Janus TMD heterobilayers.}
   \label{fig:BL_born}
\end{figure} 

The BEC tensors in the considered Janus TMD heterobilayers are dominated by their in-plane diagonal components (Figure~\ref{fig:BL_born}), while the out-of-plane terms remain close to zero over the entire strain range, and all off-diagonal elements vanish. The acoustic sum rule is satisfied with negligible numerical deviation, confirming the consistency and convergence of our calculations.
 A striking feature of these systems is the sign profile of the BECs, where the transition-metal atoms display negative values while the chalcogen species exhibit positive values (Figure~\ref{fig:BL_born}). This counterintuitive sign reversal~\cite{Pike2017} is consistent with trends reported for conventional TMD monolayers\cite{Pike2017,Yumnam2018}. Physically, this behavior does not mirror the static oxidation states of isolated metallic and chalcogen atoms, but it is caused by an electronic charge redistribution due to hybridization between the antibonding $d$-orbitals of the transition-metal species and chalcogen $p$-states  near the Fermi level~\cite{Pike2017,Yumnam2018}. The out-of-plane BEC components of the Janus heterobilayers are substantially suppressed, with a magnitude that is almost half of the corresponding in-plane values reported for pristine monolayer counterparts \cite{Pike2017}. 
 
 At zero strain, the BECs of Mo and W atoms (Figure~\ref{fig:BL_born}) lie approximately between those of their parent monolayers  (MoS$_2$/MoSe$_2$ and WS$_2$/WSe$_2$, respectively)\cite{Cheng2018,Sohier2016}, reflecting the mixed chemical environment of the asymmetric Janus structures. Under biaxial strain, all bilayers show a monotonic trend: compressive strain enhances the magnitude of the in-plane BEC components ($Z^*_{\parallel}$), making them more negative for the transition metals and more positive for chalcogens, whereas tensile strain reduces their absolute values. Although the BECs of the considered Janus TMD heterobilayers remain smaller than those typically observed in highly polar materials, such as perovskite oxides\cite{Ghosez1995,Detraux1997}, they are sufficiently large to mediate a robust electromechanical coupling, consistent with their piezoelectric response calculated herein.

\subsubsection{Effective Masses}
Modulating the electronic band dispersion via strain offers a straightforward pathway to tailor carrier transport dynamics and maximize electromechanical efficiency in Janus TMD heterobilayers. Quantifying band curvature is therefore essential, as their local dispersion directly influences carrier mobility.
Within deformation-potential theory\cite{Bardeen1950}, the carrier mobility in a two-dimensional semiconductor is given by\cite{Ju2020,Xi2012}:
\begin{equation}
\mu_{2D}=\frac{2e\hbar^{3}C}{3k_{B}T|m^{*}|^{2}E_{1}^{2}},
\end{equation}
where $\hbar$, $e$, $k_B$, and $T$ denote the reduced Planck constant, the electron charge, the Boltzmann constant, and the absolute temperature, respectively; $m^{*}$ is the carrier effective mass, $E_{1}$ is the deformation potential constant, and $C$ is the in-plane stretching modulus. 
The strain-induced modifications to the band curvature are quantified from the geometric average along the $x$- and $y$-directions:
\begin{equation}
m^{*}=\sqrt{m_x^{*}m_y^{*}}.
\end{equation}

\begin{figure}[t]
    \centering
    \includegraphics[width=0.9\linewidth]{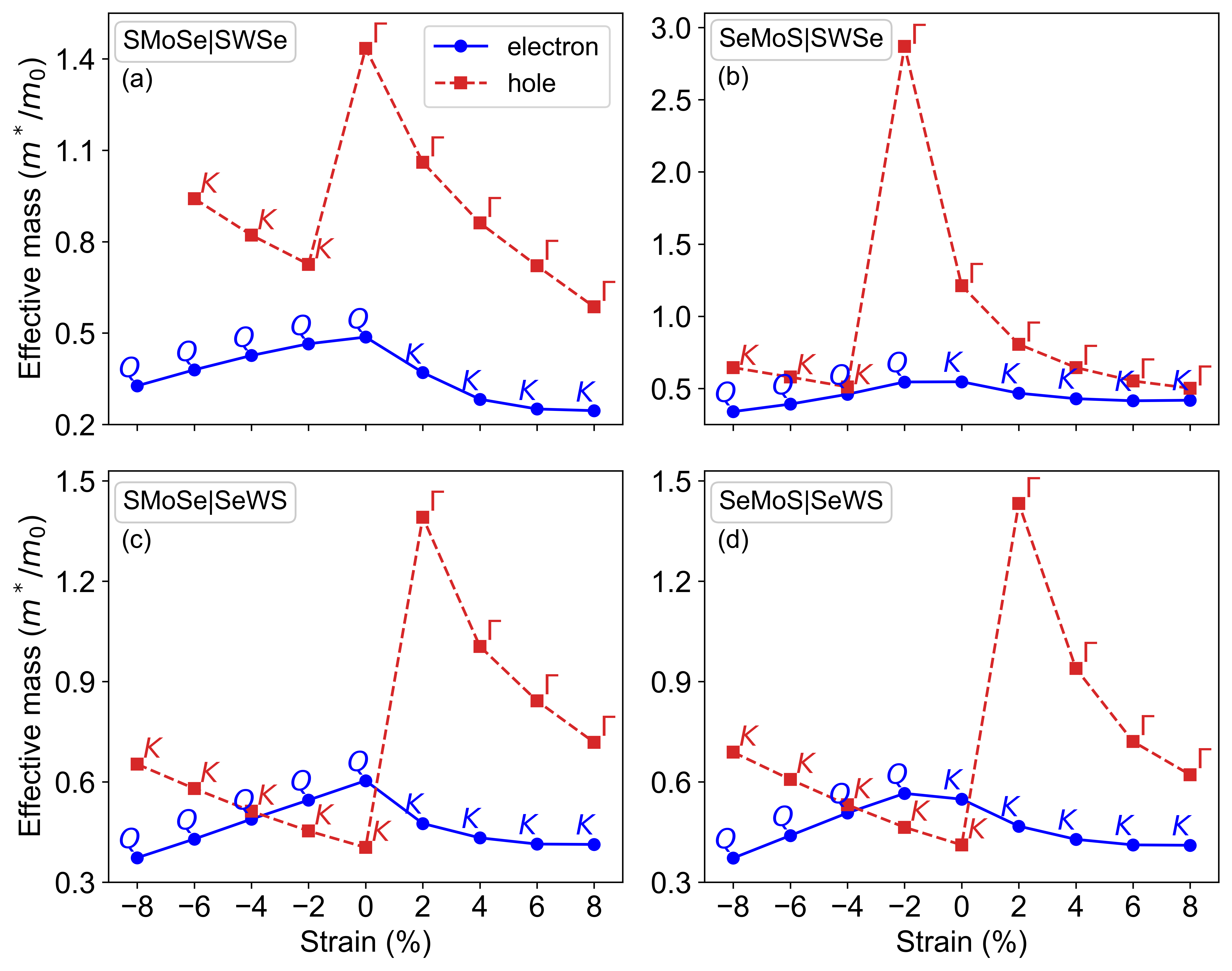}
    \caption{Electron and hole effective masses of the considered Janus heterobilayers as a function of biaxial strain. The labels of the data points indicate the valley positions.   }
    \label{fig:BL_mass}
\end{figure}

All four Janus TMD heterobilayers exhibit a consistent trend (Figure~\ref{fig:BL_mass}): the electron effective masses ($m_e^*$) remain relatively small and vary only moderately with strain, typically within the $\sim$0.3–0.6~m$_0$ range. Under compression, $m_e^*$ gradually increases, reaching a shallow maximum near the equilibrium-to-low-compression regime ($0\%$ to $-2\%$), and subsequently decreases under higher tensile strain. This variation reflects a strain-induced reconstruction of the conduction band edge, where the CBM shifts between the \textit{Q}- and the \textit{K}-valleys (Figure~\ref{fig:BL_bands}). Since these two competing valleys possess remarkably similar band curvatures, the resulting electron effective mass is highly resilient against moderate lattice deformations.

The hole effective masses ($m_h^*$) exhibit, in contrast, a much more pronounced sensitivity on biaxial strain. In the compressive strain region, $m_h^*$ remains moderate and is mainly determined by the highly dispersive \textit{K}-valley. Applying tensile strain, however, $m_h^*$ grows substantially in all heterobilayers, driven by the VBM shifting  to the Brillouin zone center. Since the $\Gamma$-valley is characterized by a significantly flatter dispersion than the \textit{K}-valley, the strain-induced valley transition triggers an abrupt enhancement of $m_h^*$, peaking at 2\% for most systems. An anomalously large value of $\sim$2.9~m$_0$ is reached in the equilibrium SeMoS|SWSe configuration (Figure~\ref{fig:BL_mass}b, -2\% strain). Under higher tensile strain, the hole masses gradually decrease as the $\Gamma$-valley disperses, although they remain consistently larger than their electron counterparts. 

Overall, biaxial strain appears to modulate the valence-band dispersion much more pronouncedly than the conduction-band dispersion. These strain-driven valley transitions among the \textit{K}-, \textit{Q}-, and $\Gamma$-points (Figure~S6, S7, S8) produce significant variations in the carrier effective masses, thereby strongly influencing carrier mobility\cite{Datye2022,Yang2024} and inducing anisotropic transport in Janus TMD heterobilayers. 

\section{Discussion and Conclusions}
 The Janus TMD heterobilayers investigated in this work exhibit a rich spectrum of electromechanical, electronic, and piezoelectric properties under biaxial strain. Driven by structural asymmetry, intrinsic electrostatic fields, interlayer coupling, and complex valley topology, these systems diverge significantly from their conventional TMD counterparts.
 
 Their mechanical response to biaxial strain is highly anisotropic along the out-of-plane axis, driven primarily by the mismatch in the chalcogen atomic radii. This structural deformation alters the macroscopic out-of-plane dipole moment $\mu_{\text{ML}}$. The monotonic increase of $\mu$ under tensile strain in the self-standing Janus monolayers, is a consequence of stretching the polar covalent bonds, which increases the spatial separation between the centers of positive (metal) and negative (chalcogen) charge. However, the transition to a non-monotonic, dome-shaped trend in the bilayers highlights an intense, non-additive electrostatic screening effect. As a result, the net dipole moment in the bilayers ($\mu_{\text{BL}}$) is not a mere linear combination of constituent monolayer dipoles; rather, it is mediated by a dynamic interlayer charge redistribution that can be modeled as:
 \begin{equation}
     \mu_{\text{BL}} = \mu_{\text{MoSSe}}(\varepsilon) + \mu_{\text{WSSe}}(\varepsilon) + \Delta \mu_{\text{int}}(\varepsilon). 
 \end{equation}
Here, $\Delta \mu_{\text{int}}(\varepsilon)$ quantifies how the strain-modulated interlayer distance $d$ modifies the interfacial  capacitance, generating a microscopic charge transfer of up to $\sim 0.02$ $e$. This interlayer charge redistribution screens the internal electric field under compression and modifies the band alignment under tension, creating the characteristic dome-shaped profile. Notably, the anti-parallel configuration SeMoS|SWSe undergoes a massive dipole cancellation, yet maintaining a robust built-in interfacial field that significantly shifts upward the absolute CBM and VBM positions relative to the vacuum level (Figure \ref{fig:BL_offset}). This behavior provides a valuable design rule for engineering the ultra-shallow CBM positions required for high-efficiency electron injection or specific photocatalytic redox reactions.\cite{Cant2014,DiLiberto2021,Kavan2019}  

Another key finding of this work is the remarkable  stability of their indirect band gap across a wide, technologically relevant strain window ($-6\%$ to $2\%$) in all systems except SeMoS|SeWS. This behavior stands in stark contrast to conventional homo- and heterobilayers like $\text{MoS}_2$ or $\text{WS}_2$, which undergo a rapid indirect-to-direct transition even under moderate tensile strain. This stability is rooted in the large energy splitting induced by the cooperative effects of intrinsic electric fields and heterometal band offsets. The conduction \textit{Q}-valley acts as an energetic anchor, remaining nearly rigid (up to variations of only $\sim 400\text{ meV}$) due to its highly localized, non-bonding atomic orbital character. Meanwhile, the \textit{K}-valley shifts significantly by nearly $2\text{ eV}$ across the same strain range. The wide energy separation between these valleys at equilibrium prevents the \textit{K}-valley from overtaking the \textit{Q}-valley under low-to-moderate strain.

Concurrently, biaxial strain provides an effective mechanism for tuning the carrier transport properties through modifications of the band dispersion/curvature. The relative insensitivity of the electron effective masses indicates that the conduction-band states preserve a highly dispersive character, leading therefore to large mobilities, even across strain-induced \textit{Q}- to \textit{K}-valley transitions. In contrast, the pronounced enhancement of the hole effective masses in the tensile strain regime originates from the shift of the VBM to the flat $\Gamma$-valley, which substantially reduces hole mobility. This asymmetric dependence of  electron and hole responses suggests that strain can be used to selectively engineer specific transport channels, particularly for isolating electron-dominated or highly anisotropic conduction paths.

At the atomic scale, the coupling between mechanical work and electrical polarization is clarified by the Born effective charges and piezoelectric responses. The anomalous sign profile of the BECs, where the transition-metal atoms display negative charges and chalcogens assume positive values, is a fingerprint of dynamic charge transfer due to intrinsic out-of-plane polarity in Janus TMDs\cite{Pike2017}, further tuned by lattice distortion. Under compression, bond shortening enhances $d-p$ orbital hybridization, forcing electronic charge back into the anti-bonding states and maximizing the dynamic charge response. On a macroscopic scale, the in-plane piezoelectric coefficient proves to be a robust, configuration-independent parameter. Being constrained by the basal plane's threefold rotational symmetry ($C_{3v}$), the intralayer atomic displacements dominate over long-range interlayer interactions. In contrast, the shear piezoelectric coefficient is highly sensitive to the stacking configuration. In the presence of antiparallel dipole alignments, the shear response is suppressed through cancellation of opposing out-of-plane dipoles, while parallel dipole alignments break macroscopic inversion symmetry, unlocking a large, prominent net shear polarization. 

In view of the diverse characteristics discussed above, these Janus TMD heterobilayers appear particularly promising for specialized electronic applications. The stable-gap configurations (SMoSe|SWSe, SeMoS|SWSe, SMoSe|SeWS) are ideally suited for flexible, wearable optoelectronics and photovoltaics, ensuring consistent performance under physical deformation. In contrast, the sensitive SeMoS|SeWS stack is a premier candidate for nano-electromechanical strain sensors and logic switches, where even small compressive strain triggers electronic switching events. Finally, the shear piezoelectricity appearing in parallel-aligned symmetric interfaces offers a suitable platform for high-precision energy harvesters and smart actuators capable of converting mechanical vibrations into directional electrical power. By engineering the interfacial chemistry, these materials can be transformed from passive components into active, responsive elements tailored to the operational environment. 
 
In summary, using first-principles calculations, we have demonstrated that the interfacial chemistry of Janus TMD heterobilayers is an effective design parameter to overcome the strain sensitivity inherent to conventional TMDs. By manipulating the chalcogen stacking sequence at the interface, we achieved a high degree of control over the electronic band structures, carrier effective masses, and piezoelectric tensors across a wide biaxial strain window. The intrinsic out-of-plane dipoles and heterometal band offsets act as stabilizing mechanisms, preserving the indirect band gap and suppressing destructive transitions common to conventional TMDs. Furthermore, the ability to toggle the shear piezoelectric coefficient on and off through interfacial symmetry breaking introduces a novel and versatile degree of freedom for electromechanical device miniaturization. Ultimately, interlayer electrostatics and dynamic charge redistribution establish a robust physical framework for achieving strain-durable functionality, positioning Janus TMD heterobilayers as prime candidates for next-generation flexible optoelectronic, valleytronic, and nano-electromechanical applications.

\section{Computational Methods}
All first-principles calculations presented in this work were performed using plane-wave density-functional theory~\cite{Hohenberg1964,Kohn1965} as implemented in the Vienna Ab initio Simulation Package (VASP).~\cite{Kresse1996} Electron-ion interactions were described via the projector augmented wave method ~\cite{Kresse1999,blochl1994}, adopting a kinetic energy cutoff of 550~eV. The exchange-correlation potential was treated within the generalized gradient approximation parametrized by Perdew, Burke, and Ernzerhof~\cite{Perdew1996} and augmented by the Grimme-D3 semi-empirical correction to capture long-range dispersion forces\cite{Grimme2010,Grimme2011}.  

To remove spurious interactions between periodic images in the out-of-plane direction, a vacuum layer exceeding 20~\AA{} was included along \textit{z}. Additionally, a dipole correction was applied to cancel artificial electrostatic interactions stemming from the intrinsic polarity of the Janus structures. The Brillouin zone was sampled with a $\Gamma$-centered $15 \times 15 \times 1$ \textbf{k}-point grid. Spin-orbit coupling was included in all calculations. For structural relaxation, the total energy was converged below $10^{-6}$~eV, and the Hellmann-Feynman forces were minimized below 0.01 eV/\AA{} per atom.

The macroscopic piezoelectric stress tensor components and the microscopic Born effective charges were calculated using density functional perturbation theory \cite{baroni2001,wu2005}. Post-processing and data analysis were carried out with the VASPKIT~\cite{WANG2021} tool, \texttt{pymatgen}\cite{ONG2013}  and in-house-produced \texttt{Python} scripts\cite{torkashvand2026}.


\begin{acknowledgement}
This work was partly funded by the German Research Foundation, Project numbers 398816777 (CRC 1375, sub-project A8) and 547611111 (WHAT-A-TWIST). Computational resources were provided by the Friedrich-Schiller University Jena (cluster ARA). 

\end{acknowledgement}

\begin{suppinfo}
The following files are available free of charge.
\begin{itemize}
 \item {si.pdf: The supporting information includes:Janus monolayer band-structure plots, Formation energy and interlayer binding energy of the Janus bilayers, Various high-symmetry k-point gaps and effective masses for the Janus bilayers.} 
 
\end{itemize}

\end{suppinfo}
\vspace{0.5 cm}

\noindent\textbf{Data Availability Statement}\\
Input and output files of the ab initio calculations performed in this work are available free of charge in Zenodo \href{https://doi.org/10.5281/zenodo.20449173}{DOI: 10.5281/zenodo.20449173
} [record: 20449173]. 

\noindent\textbf{Notes}\\
The authors declare no competing financial interest.

\bibliography{main}

\end{document}


\newpage

\section{Electronic Properties of Janus Monolayers}
\begin{figure}
    \centering
    \includegraphics[width=0.995\linewidth]{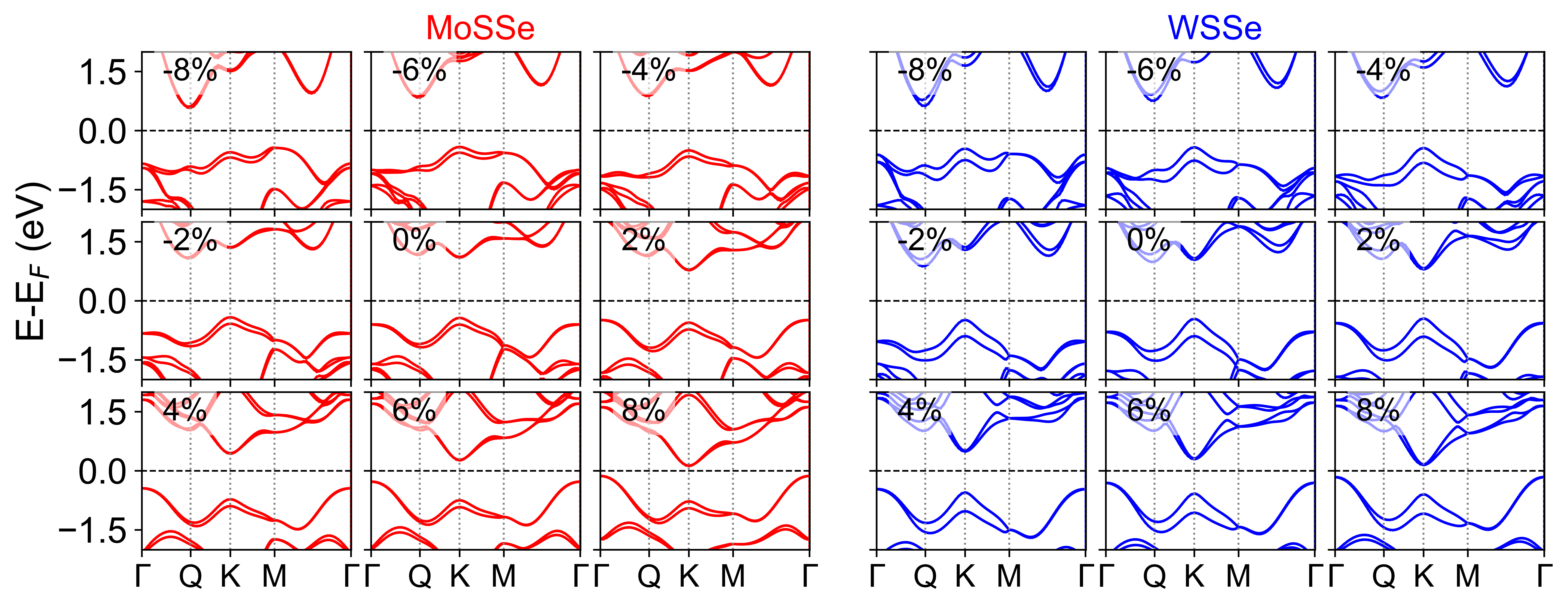}
    \caption{Band structure of Janus MoSSe and WSSe monolayers with applied biaxial strain.}
    \label{fig:BL_bands_shift}
\end{figure} 

\section{Structural Properties of Janus Bilayers}
\subsection{Formation and interlayer binding energy of Janus bilayers}
\begin{table}[h]
\centering
\renewcommand{\arraystretch}{1.15}
\setlength{\tabcolsep}{4pt}

\resizebox{\linewidth}{!}{%
\begin{tabular}{c|cccc|cccc}
\hline
 & \multicolumn{4}{c|}{Binding energy (eV)} & \multicolumn{4}{c}{Formation energy (eV)} \\
\hline
Strain (\%) 
& SMoSe|SWSe & SeMoS|SWSe & SMoSe|SeWS & SeMoS|SeWS 
& SMoSe|SWSe & SeMoS|SWSe & SMoSe|SeWS & SeMoS|SeWS \\
\hline
-8 & -0.289 & -0.263 & -0.301 & -0.284 & -5.764 & -5.760 & -5.766 & -5.763 \\
-6 & -0.284 & -0.259 & -0.295 & -0.279 & -5.869 & -5.865 & -5.871 & -5.868 \\
-4 & -0.279 & -0.256 & -0.289 & -0.274 & -5.939 & -5.935 & -5.940 & -5.938 \\
-2 & -0.275 & -0.253 & -0.283 & -0.269 & -5.978 & -5.974 & -5.979 & -5.977 \\
 0 & -0.271 & -0.249 & -0.279 & -0.265 & -5.990 & -5.987 & -5.992 & -5.989 \\
 2 & -0.267 & -0.246 & -0.275 & -0.261 & -5.979 & -5.976 & -5.981 & -5.978 \\
 4 & -0.263 & -0.243 & -0.272 & -0.258 & -5.947 & -5.944 & -5.949 & -5.946 \\
 6 & -0.260 & -0.240 & -0.269 & -0.258 & -5.897 & -5.893 & -5.898 & -5.896 \\
 8 & -0.259 & -0.240 & -0.266 & -0.260 & -5.830 & -5.827 & -5.831 & -5.830 \\
\hline
\end{tabular}%
}

\caption{Comparison of interlayer binding energy and formation energy (in eV) for four bilayer configurations under strain ranging from $-8\%$ to $8\%$.}
\end{table}

\newpage
\section{Electronic Properties of Janus Bilayers}
\subsection{Individual atom Bader charges in bilayers}
\begin{figure}[h]
    \centering
    \includegraphics[width=0.9\linewidth]{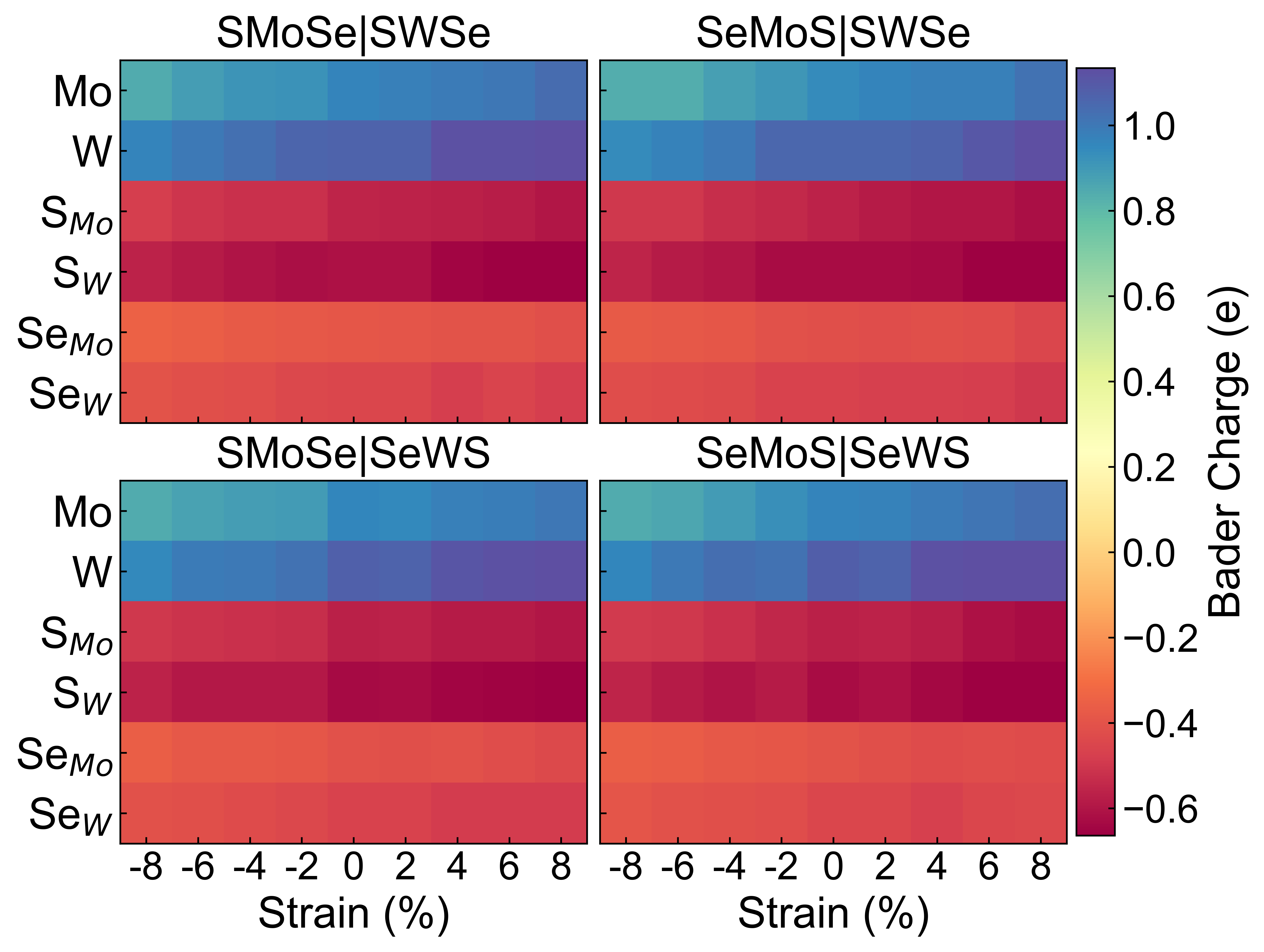}
    \caption{Trends of the individual atom Bader charges variation  with strain for all bilayers.}
    \label{fig:BL_bands_shift}
\end{figure} 

\newpage

\subsection{Energy Levels in Janus Heterobilayers}
\begin{figure}[h]
    \centering
    \includegraphics[width=0.95\linewidth]{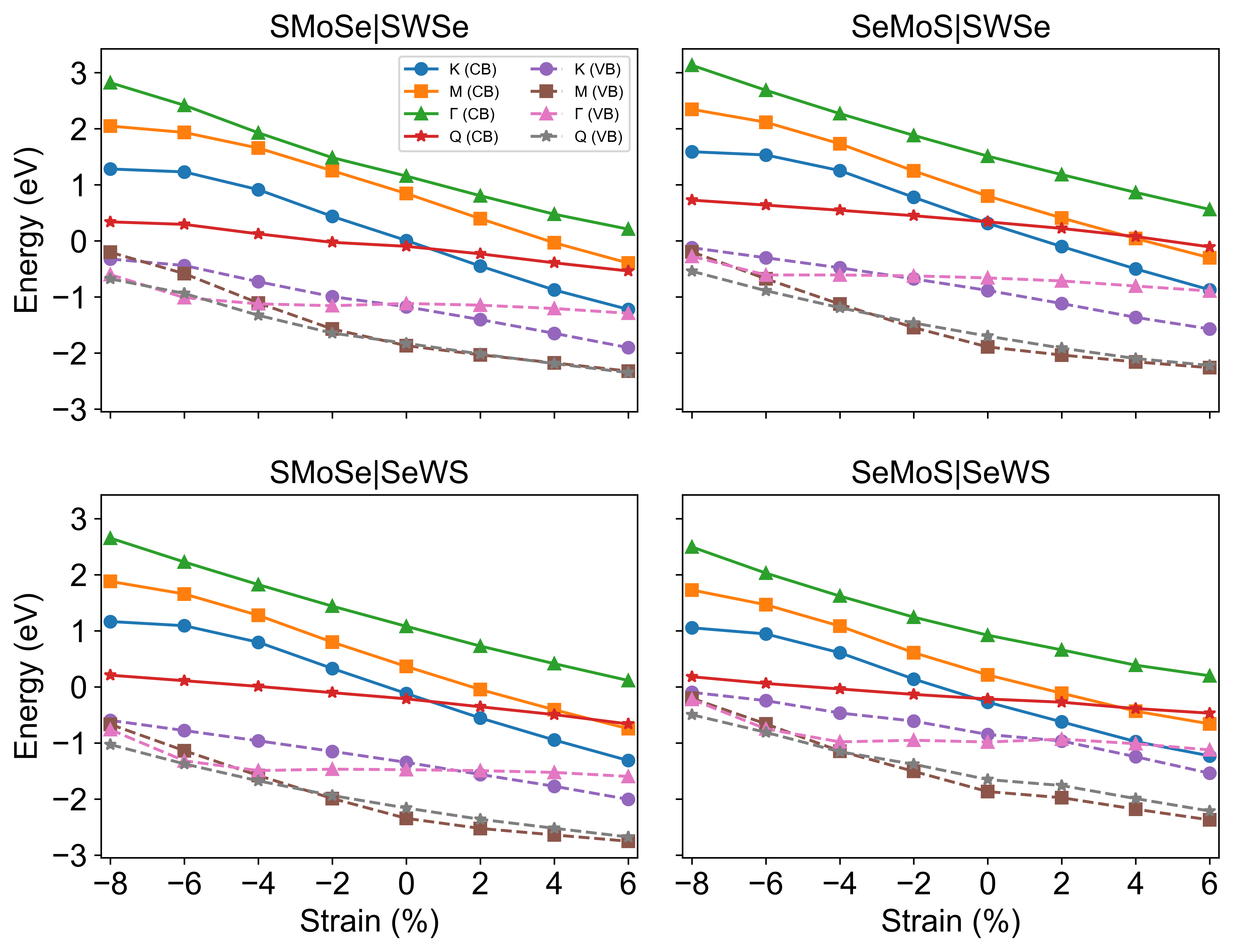}
    \caption{Strain-dependent energies of the highest valence band (VB) and lowest conduction band (CB) at different high-symmetry points in all considered Janus heterobilayers.}
    \label{fig:BL_bands_shift}
\end{figure} 
\newpage

\subsection{Various high-symmetry points direct and indirect energy gaps}
\begin{figure}
    \centering
    \includegraphics[width=0.9\linewidth]{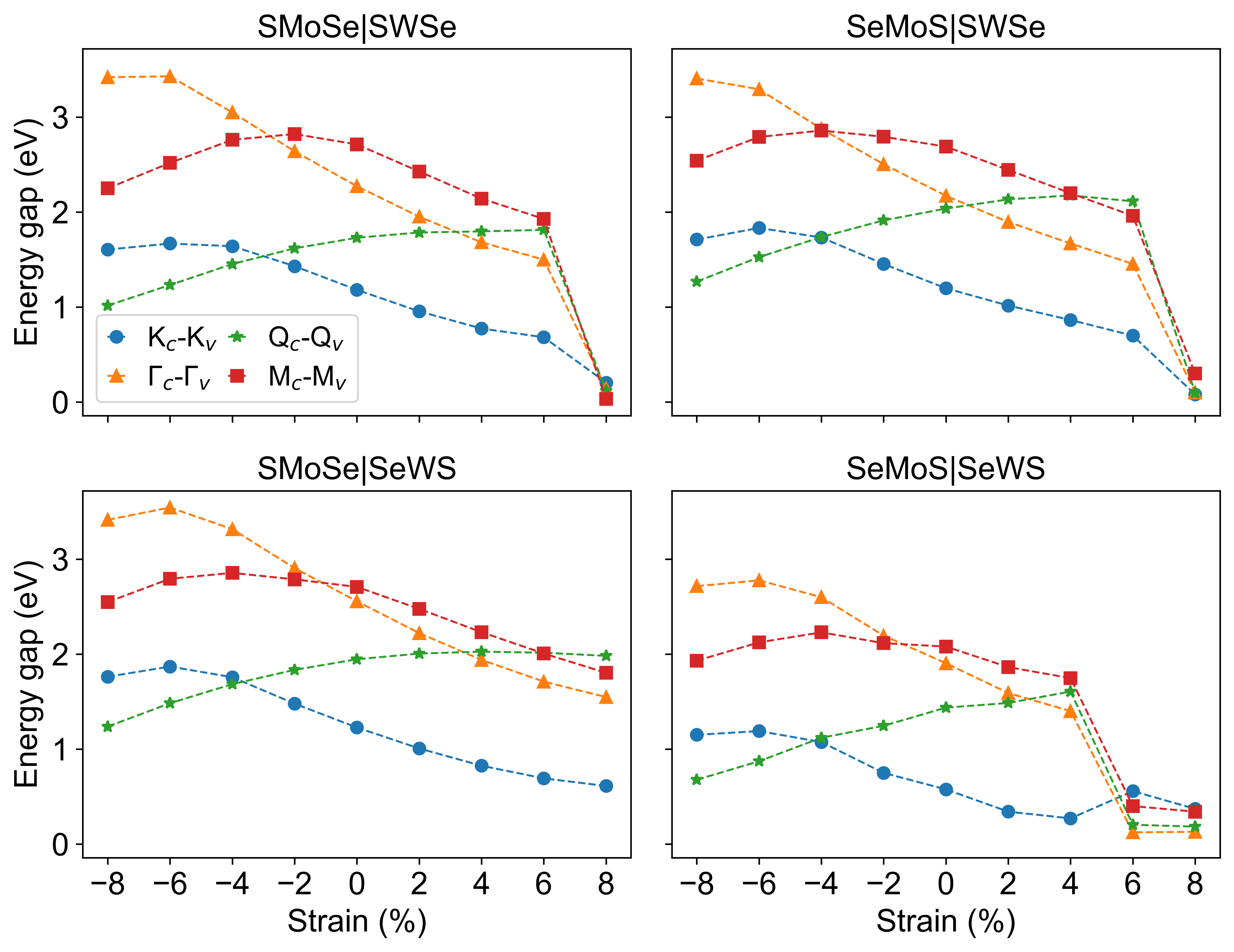}
 
    \caption{Direct  energy gaps at different high-symmetry points as a function of biaxial strain for all bilayers.}
    \label{fig:BL_gaps_direct}
\end{figure}
\begin{figure}
    \centering

    \includegraphics[width=0.9\linewidth]{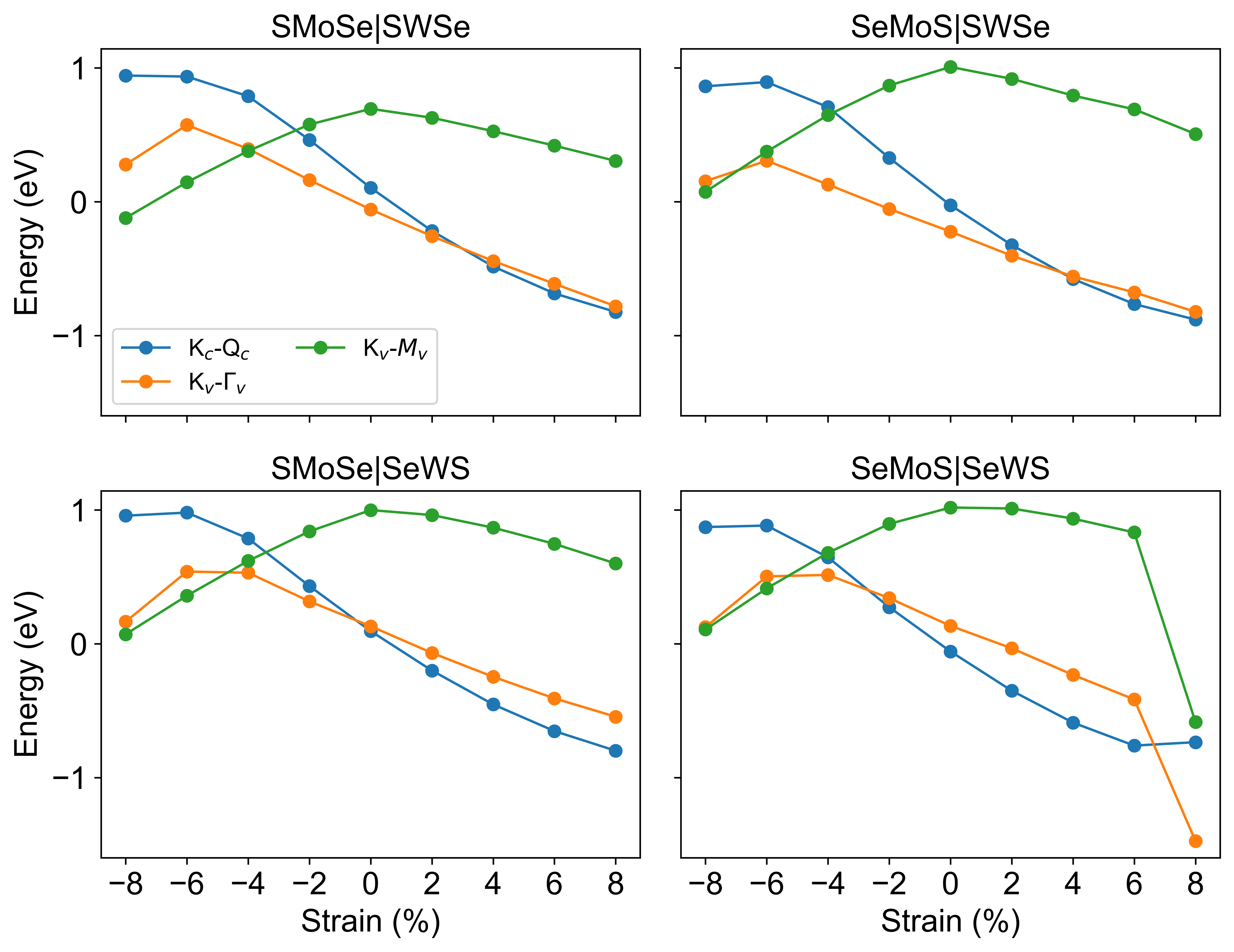}

    \caption{Indirect energy gaps at different high-symmetry points as a function of biaxial strain for all bilayers.}
    \label{fig:BL_gaps_direct}
\end{figure}

\newpage
\subsection{Various high-symmetry point effective masses}
\begin{figure}
    \centering

    \includegraphics[width=0.905\linewidth]{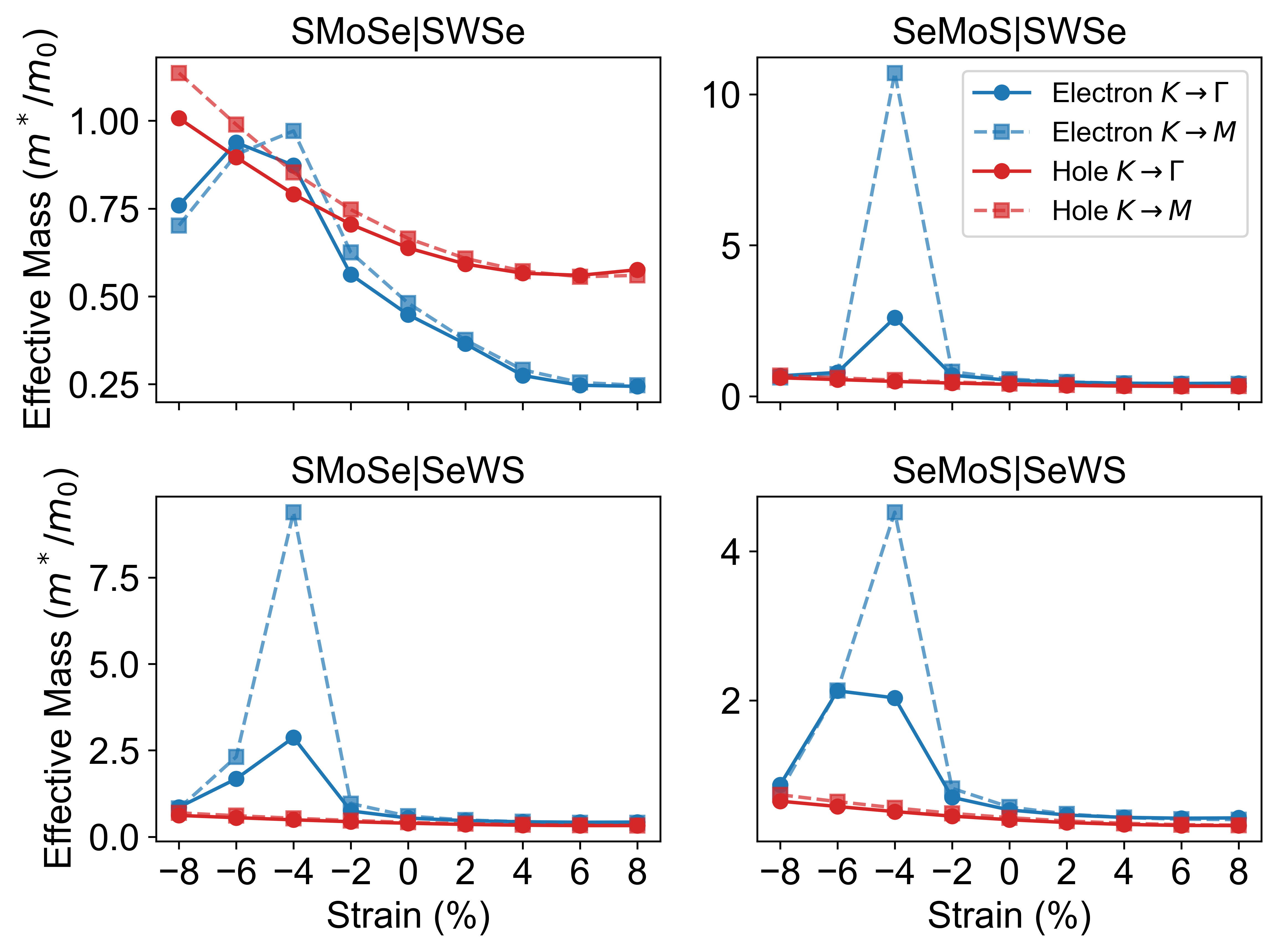}

    \caption{Electron and hole effective masses  at K-point as a function of biaxial strain for all bilayers.}
    \label{fig:BL_mass}
\end{figure} 
\begin{figure}[t]
    \centering

    \includegraphics[width=0.905\linewidth]{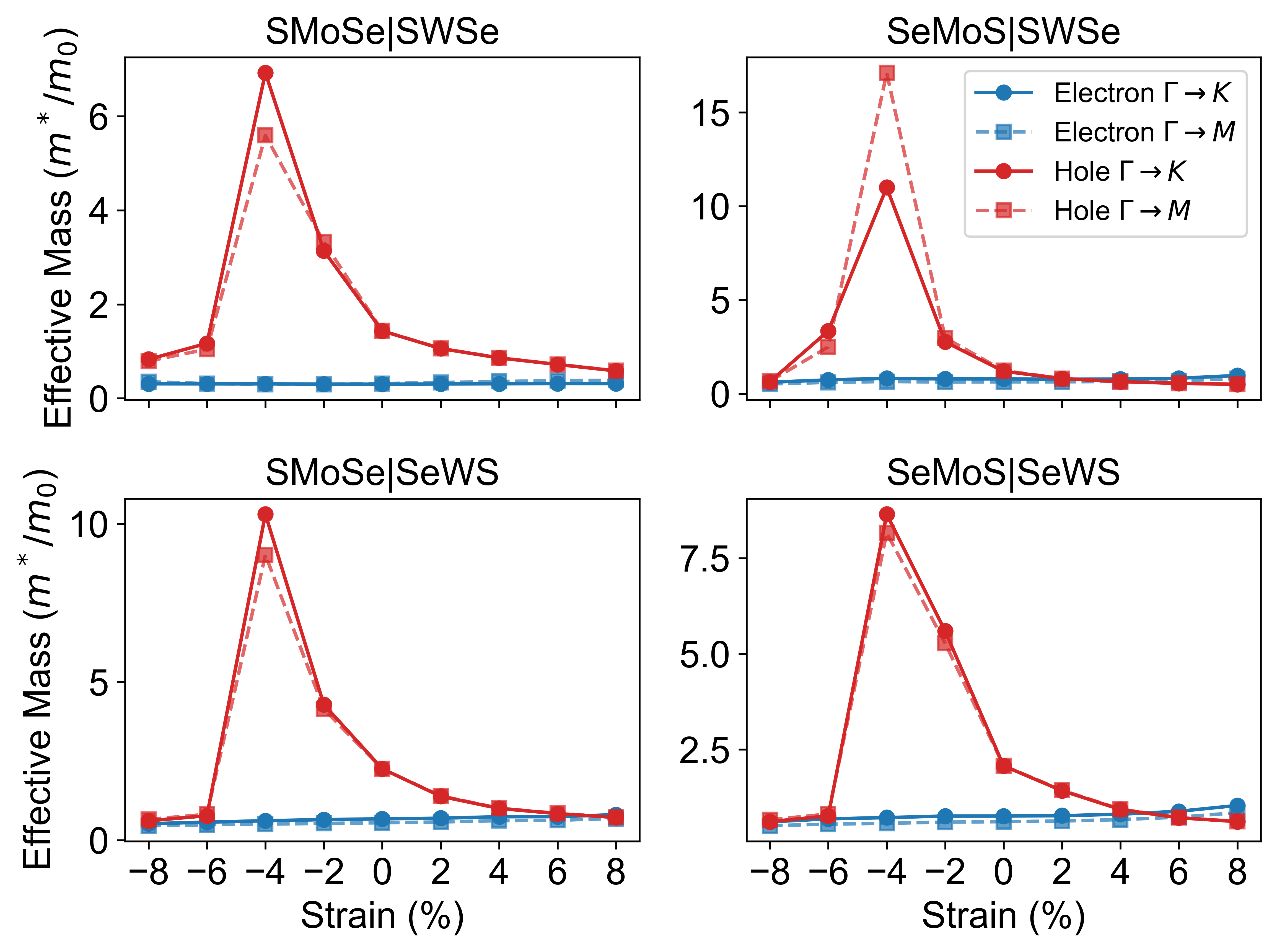}
    
    \caption{Electron and hole effective masses  at $\Gamma$-point as a function of biaxial strain for all bilayers.}
    \label{fig:BL_mass}
\end{figure} 

\begin{figure}[t]
    \centering

    \includegraphics[width=0.905\linewidth]{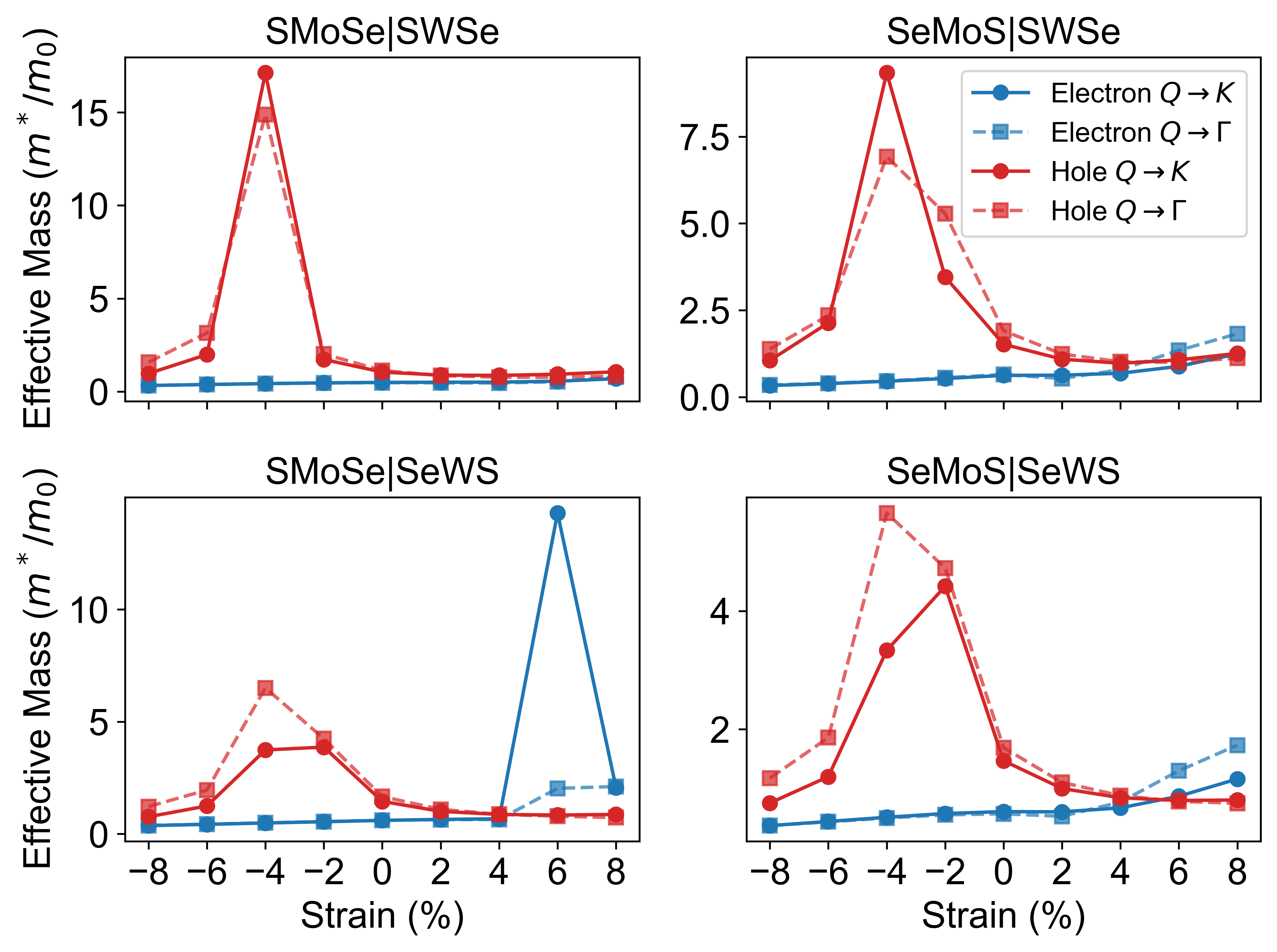}
    \caption{Electron and hole effective masses  at Q-point as a function of biaxial strain for all bilayers..}
    \label{fig:BL_mass}
\end{figure} 